\documentclass[aps, prd, superscriptaddress, nofootinbib, preprintnumbers]{revtex4}
\usepackage{amsmath}
\usepackage{bbold}
\usepackage{graphicx}
\usepackage{color}

\newcommand{\chushi}[1]{}

\newcommand{\beq}{\begin{eqnarray}}
\newcommand{\eeq}{\end{eqnarray}}
\begin{document}
\title{
\begin{flushright}
\begin{minipage}{0.2\linewidth}
\normalsize
MITP/18-112 \\*[50pt]
\end{minipage}
\end{flushright}
\bf 
Gravitational Waves from Walking Technicolor
 \vskip 0.5cm
}

\author{Kohtaroh~Miura}
\thanks{{\tt kohmiura@uni-mainz.de}}
\affiliation{Helmholtz-Institut Mainz, Johannes Gutenberg-Universit\"{a}t Mainz, D-55099 Mainz, Germany}
\affiliation{Kobayashi-Maskawa Institute for the Origin of Particles and the Universe,\\
Nagoya University, Nagoya 464-8602, Japan}

\author{Hiroshi~Ohki}
\thanks{{\tt hohki@asuka.phys.nara-wu.ac.jp}}
\affiliation{Department of Physics, Nara Women's University, Nara 630-8506, Japan}

\author{Saeko~Otani}
\thanks{{\tt otani@asuka.phys.nara-wu.ac.jp}}
\affiliation{Department of Physics, Nara Women's University, Nara 630-8506, Japan}

\author{Koichi~Yamawaki}
\thanks{{\tt yamawaki@kmi.nagoya-u.ac.jp}}
\affiliation{Kobayashi-Maskawa Institute for the Origin of Particles and the Universe,\\
Nagoya University, Nagoya 464-8602, Japan}

\begin{abstract}
We study gravitational waves from the first-order electroweak phase transition in the $SU(N_c)$ 
gauge theory with $N_f/N_c\gg 1$ (``large $N_f$ QCD'') 
as a candidate for the walking technicolor, which is modeled by the $U(N_f)\times U(N_f)$ linear sigma model
with classical scale symmetry (without mass term), particularly for $N_f=8$ (``one-family model'').
This  model  exhibits
spontaneous breaking of the scale symmetry as well as the $U(N_f)\times U(N_f)$ 
radiatively through the Coleman-Weinberg mechanism $\grave{a}$  la Gildener-Weinberg, 
thus giving rise to a light pseudo dilaton 
(technidilaton)  to be identified with the 125 GeV Higgs.
This model possess a strong first-order electroweak phase transition 
due to the resultant Coleman-Weinberg type potential. 
We estimate the bubble nucleation that exhibits an ultra supercooling 
and then  the signal for a stochastic gravitational wave produced via
the  strong first-order electroweak phase transition. 
We show that the amplitude 
can be reached to the expected sensitivities of the LISA.
\end{abstract}

\maketitle

\section{Introduction}

The origin of mass is one of the most important issues in 
particle physics.
In the standard model (SM), 
the parameters of the Higgs boson mass and the electroweak symmetry breaking scale 
are all free parameters, 
so that the dynamical origin of mass is a challenging issue.
One attractive model beyond the standard model 
is the walking technicolor, 
in which the electroweak symmetry is dynamically broken 
due to 
approximately scale-invariant 
strong gauge dynamics (i.e. ladder Schwinger-Dyson (SD) equation) with large anomalous dimension $\gamma_m \simeq 1$~\cite{Yamawaki:1985zg,
Bando:1986bg},\footnote{The walking technicolor was subsequently studied \cite{Akiba:1985rr,Appelquist:1986an} without notion of the scale invariance/dilaton and anomalous dimension,
 along the line of the earlier numerical study \cite{Holdom:1984sk}.
}
  in contrast to the original technicolor~\cite{Weinberg:1975gm, Susskind:1978ms} as a simple scale-up of the QCD. 
Due to 
the spontaneously broken approximate scale invariance, the walking technicolor model predicts~\cite{Yamawaki:1985zg,
Bando:1986bg} 
a pseudodilaton (``technidilaton''),  a pseudo Nambu-Goldstone (NG) boson of the spontaneously broken approximate scale symmetry, which could 
 be identified with 
the Higgs boson with mass $125$ GeV (see \cite{Matsuzaki:2015sya} and references cited therein).

The most popular candidate dynamics for 
the  walking technicolor 
is the  ``large $N_f$ QCD'', 
the $SU(N_c)$ gauge theory with $N_f (\gg N_c) $ fermions in the fundamental representation,
where the theory has the Caswell-Banks-Zaks (CBZ) infrared 
fixed point~\cite{Caswell:1974gg,Banks:1981nn}  for some larger $N_f$ (conformal window),  
thereby becoming scale invariant at the fixed point.
It is then expected  that a walking gauge theory is realized  for $N_f$
just below the conformal window, 
the phase boundary of which 
is roughly estimated to be 
$N_f/N_c
\lesssim 4$ \cite{Appelquist:1996dq},
based on 
the two-loop beta function 
combined with the ladder SD 
equation, which is  justified at least qualitatively for the large $N_f$ QCD  in the ``anti-Veneziano limit'' $N_c\rightarrow \infty$ with $N_f/N_c ={\rm fixed} \gg 1$ \cite{Matsuzaki:2015sya} (cf. Veneziano limit with $N_f/N_c \ll 1$).
In the walking phase 
the (approximate) scale symmetry as well as the chiral symmetry   is spontaneously broken, thus we expect existence of 
the flavor-singlet scalar meson, 
$\sigma$, as light enough to be identified with the pseudodilaton.

Recently the lattice simulations of large $N_f$ QCD for $N_c=3$
have been widely studied 
to find 
such a walking behavior. Among others  it was found~\cite{Aoki:2013xza,Appelquist:2014zsa,Hasenfratz:2014rna} 
that $N_f=8, N_c=3$ QCD 
has a walking behavior with signals of the spontaneously broken phase 
and the approximate scale symmetry (non-universal 
hyperscaling with $\gamma_m \simeq 1$).
More interestingly, 
a light $\sigma$ meson was in fact discovered  
in the  $N_f=8, N_c=3$  QCD on the lattice 
~\cite{Aoki:2014oha,
Aoki:2016wnc,Appelquist:2016viq,Appelquist:2018yqe}.
While the results 
have been obtained 
in the presence of the non-zero fermion mass $m_f$, 
in contrast to 
the walking technicolor model 
defined 
at vanishing fermion mass, 
it encourages 
us to consider a possibility of the composite dilaton (=Higgs/techinidilaton)  
as a light bound state of the flavor singlet scalar 
in the large $N_f$ QCD. 

The
$N_f=8$ QCD is also attractive from the 
point  of view of the model building, 
since 
the 8 flavors of the fermion in this model can be assigned 
as the four electroweak doublets 
in the one-family technicolor model, 
which is a model 
naturally fit in 
the extended technicolor (ETC) model 
to explain the dynamical origin of all  the masses including the quarks and leptons~\cite{Farhi:1980xs}.
See \cite{Matsuzaki:2015sya} for  the solution of the old problems such as the flavor-changing neutral current problems 
and the oblique corrections (S parameters, etc.), 
and the explicit breaking mass of technipions (pseudo NG bosons of larger chiral symmetry), etc. 
in the one-family walking technicolor model.

Another interesting feature in large $N_f$ QCD is 
its thermodynamic properties.
As 
argued by Pisarski and Wilczek~\cite{Pisarski:1983ms}, 
in the QCD for 
$N_f>2$ 
the finite temperature phase transition  
should be of first-order. If it is the case,  the large $N_f$ QCD as a walking technicolor 
should be quite attractive from the view of physics beyond the SM,
since the SM which, having the same symmetry structure as that of the $N_f=2$ QCD,
is believed to have the strength of the electroweak phase transition not strong enough.  
Actually it is a key condition to explain the universe's matter dominance  
in the electroweak baryogenesis scenario.
Thus a possibility for the strong first-order electroweak phase transition  
should be a key ingredient for the walking technicolor as a large $N_f$ QCD to account for 
the origin of the universe's matter dominance.

In this paper
we examine if the transition of the finite temperature electroweak phase transition 
in the early universe 
is of the first-order 
in the  large $N_f$  QCD with $N_f=8$ as a walking technicolor model. 
If it is the case, the gravitational wave signal will be a good probe to the first-order 
phase transition in the early universe~\cite{Witten:1984rs}.

In fact such a gravitational wave can be detected by the future satellite observations 
such as the LISA~\cite{Seoane:2013qna,Audley} 
and DECIGO~\cite{Seto:2001qf,Sato:2017dkf} 
whose sensitivity will reach the magnitude of the expected electroweak phase transition.
Based on these recent developments on the gravitational wave signals, 
we here examine the electroweak bubble nucleation in thermal phase transition. 
Since fully nonperturbative analyses of the large $N_f$ QCD such as the lattice studies are presently not available for this purpose, 
we need to adopt some effective theory approach.

An effective theory of the large $N_f$ QCD as a candidate for the walking technicolor has been given by the (approximately) scale-invariant 
nonlinear sigma model, including the non-zero  fermion mass effects to be compared with the lattice data \cite{Matsuzaki:2013eva}.\footnote{ 
For subsequent studies on the candidate effective theory 
for a light 
scalar boson in the large $N_f$ QCD on the lattice, 
see Refs.~\cite{Golterman:2016lsd, Kasai:2016ifi, Appelquist:2017wcg, Hansen:2016fri} via nonlinear realization and \cite{Meurice:2017zng, Appelquist:2018tyt} on the linear sigma model.
}
 In the chiral limit $m_f=0$ which is the case for the walking technicolor, this is a 
 version of the nonlinear sigma model  consisting of $N_f^2-1$ composite pseudo NG bosons living on the manifold $G/H=SU(N_f)_L\times SU(N_f)_R/SU(N_f)_V$ plus a composite flavor-singlet scalar, $\sigma$, or the dilaton $\phi=\ln (\sigma/\langle \sigma\rangle$ ($\langle \phi\rangle=0$) with the decay constant $F_\phi=\langle \sigma \rangle$, to realize the scale symmetry nonlinearly. In addition it
possesses a small explicit breaking of the scale symmetry in the form of the Coleman-Weinberg (CW) type log potential consisting only of the dilaton (now becoming a pseudodilaton (technidilaton) identified with the 125 GeV Higgs), which reproduces the trace anomaly 
in the underlying  large $N_f$ QCD (see \cite{Matsuzaki:2015sya} and the references therein). 
 The appearance of the CW type dilaton potential is already a good signal of the first-order phase transition.

In this work, however, we adopt a more conventional systematic approach towards this problem,
based on the perturbation for the linear sigma model as an effective theory of the large $N_f$ QCD with $N_f=8$.
\footnote{
The linear sigma model has an obvious advantage over the nonlinear sigma model as to the loop calculations which are vital to the study of the gravitational waves from the phase transition due to the thermal loops. In fact the nonlinear sigma model  is subject to 
 many extra arbitrary parameters associated with the loop-induced counter terms even at one loop and zero temperature, as is well-known for the chiral Lagrangian for the hadron physics where such arbitrary parameters are actually estimated by a plenty of experimental inputs, in sharp contrast to the present case lacking such experimental inputs.
}
(For a different approach based on the holographic walking technicolor, see \cite{Chen:2017cyc}).
The model has the same symmetry as the underlying large $N_f$ QCD, namely the chiral $U(N_f)_L\times U(N_f)_R$
symmetry together with the scale symmetry (without scalar mass term) at classical level.
These symmetries are spontaneously broken as well as explicitly broken by the chiral anomaly 
and the trace anomaly at the quantum level in the underlying theory. 
Also in the linear sigma model it is possible to break the scale and the chiral symmetry spontaneously 
at  quantum level by the conventional 
CW mechanism now consistently formulated in the perturbation at one loop in a large $N_f$ version 
proposed by Gildener and Weinberg (GW) \cite{Gildener:1976ih,Paterson:1980fc}.
In the GW model, there exists a scale-invariant flat direction  along the flavor-singlet field $\sigma$
in the classical potential at some renormalization scale, say $\mu_{_{GW}}$,
which makes the perturbation fully consistent.
The flat direction actually is lifted by quantum corrections at one loop, 
so that the spontaneous chiral and scale symmetry breaking 
occurs 
in this direction, with the scale symmetry also broken explicitly as the trace anomaly due to the very loop effects. 

One striking feature in the GW mechanism is that the flavor-singlet scalar $\sigma=e^{\phi/\langle \sigma\rangle}$ becomes  
a light scalar (what they called ``scalon'') after the symmetry breaking, 
which is nothing but a 
pseudodilaton $\phi$, 
the pseudo NG  
boson of the 
spontaneously broken scale symmetry (also explicitly broken by the trace anomaly). The CW potential consists only of this pseudodilaton. 
We will 
in fact see that the CW 
potential obtained in the perturbation at the effective theory level is consistent with the aforementioned 
scale-invariant nonlinear sigma model with the CW type dilaton 
potential~\cite{Matsuzaki:2013eva} based on the nonperturbative method, 
the ladder approximation/anti-Veneziano limit, in the underlying large $N_f$ QCD.
Thus the free parameters  in our model may be estimated, the quartic couplings $<1$ 
and $\langle \sigma\rangle =F_\phi \simeq 5\,  v_{\rm EW} \simeq 1.25$ TeV ($\gg v_{\rm EW}=246$ GeV),
under matching with those of the  CW type dilaton potential in the $N_f=8$ walking technicolor~\cite{Matsuzaki:2013eva}, 
which were well fit to the  LHC data for the 125 GeV Higgs.

With the parameters so determined as bench mark values in the linear sigma model as an effective theory of the large $N_f$ QCD with $N_f=8$, we find  that 
in the corresponding walking technicolor we naturally obtain a very strong first-order phase transition, 
thanks to the nature of the GW formulation of the CW, in sharp contrast to 
the SM. 
We further find 
that this phase transition yields an ultra supercooling, from which a significant amount of the gravitational 
wave signals can be produced to be tested in future experiments such as LISA and DECIGO.

It should be noted here that 
in our model it is not necessary to introduce an elementary Higgs scalar particle 
in sharp contrast to the other composite Higgs boson scenarios 
such as dark QCD.
Recent studies of the 
gravitational wave signals from the first-order phase transition 
in composite Higgs models can be found in e.g.~\cite{Tsumura:2017knk, Marzola:2017jzl, Aoki:2017aws, Croon:2018erz, Prokopec:2018tnq}. 
In most of models, the thermal phase transition can occur in a two (or more)-dimensional 
field space, which makes the bubble nucleation analysis more complicated. 
In our model, however we provide a simple mechanism that can exhibit the electroweak symmetry breaking, 
together with a non-zero Higgs (pseudodilaton) vev.
\\

The paper is organized as follow.
In sec. II, the setup of our model will be shown,
where the dilaton potential for the singlet scalar can be obtained with a quantum effect.
It is explained how our model is dedicated to be the effective model of the walking technicolor model.
In Sec. III, we study the finite temperature effect on the dilaton potential
for which we see a strong first-order phase transition.
In Sec. IV, we investigate the bubble nucleation dynamics and the gravitational wave spectra in this model.
We also provides some discussion on the flavor non-singlet scalar mass and show the corresponding phase diagram.
Sec. V is devoted to a summary and concluding remarks.
Some details of the formulas for the external mass effects and the gravitational wave spectra
are provided in Appendix~\ref{app:ms} and \ref{app:gw}, respectively.
Throughout the paper 
all the numerical results are given with fixed $N_f=8$.
However our analytical expressions given in terms of $N_f$ are also applicable to generic $N_f$.

\section{Scale invariant linear sigma model and dilaton effective potential}

We consider a linear sigma model 
having $U(N_f)_L \times U(N_f)_R$ chiral symmetry as an effective theory of the underlying large $N_f$ QCD. 
A matrix of $N_f \times N_f$ denoted by $M_{ab}$ is an effective scalar field, 
which transforms under chiral symmetry of $U(N_f)_L \times U(N_f)_R$ as 
\begin{align}
M \to  g_L M g_R^\dagger, \quad g_L, g_R \in U(N_f).
\end{align}  
A basis of $N_f \times N_f$ Hermitian matrices $T^a (a=0, \cdots,N_f^2-1)$ is 
used to become flavor diagonal, which satisfy the following conditions as 
\begin{align}
{\rm Tr} [T^a,T^b] = 
\delta^{ab}/2.
\end{align}  
Here $T^0=\frac{{\bf 1}}{\sqrt{2N_f}} \mathbb{1}_{_{N_f\times N_f}}$ is the non traceless matrix.
The matrix field $M$ can be decomposed as
\begin{align}
M = \sum_{a=0}^{N_f^2-1} (s^a+ip^a) T^a,
\end{align}  
where $s^a$ and $p^a$ are the $N_f^2$ scalars ($0^{++}$) 
and $N_f^2$ pseudoscalars ($0^{-+}$), respectively. 
  
We consider the following renormalizable effective Lagrangian with classically scale invariance,  
\begin{align}
\label{linearmodel}
\mathcal{L} = 
{\rm Tr}[\partial_\mu M^\dagger \partial^\mu M] - V_0(M), 
\end{align}  
where $V_0$ is the tree level potential given by $U(N_f)_L \times U(N_f)_R$ invariant forms,
\begin{align}
V_0 = f_1 \left({\rm Tr}[M^\dagger M]\right)^2 + f_2 {\rm Tr}[(M^\dagger M)^2]. 
\end{align}  
Since we assume that the chiral symmetry breaks down to the diagonal subgroup $SU(N_f)_L$, 
it is natural to restrict the vev 
to its hypersphere,
\begin{align}
\langle M 
\rangle = 
T^0 \langle s^0\rangle =\frac{1}{\sqrt{2 N_f} } \mathbb{1} \cdot  \langle s^0\rangle.
\end{align} 
Following Ref.~\cite{Gildener:1976ih} 
let us briefly review the potential analysis at one loop level (see also \cite{Paterson:1980fc} for the $U(N_f)_L\times U(N_f)_R$ notation).
If we have 
\begin{align}
\label{GWcond}
f_1=-f_2/N_f,\quad  f_2>0, 
\end{align}
there is a stationary point of the potential with $V_0=0$ for non-zero 
values for $M$. 
In particular, if we restrict the field value of $M$ 
to its radial component, 
the potential has a flat direction along the radial component which can always be chosen to be $s^0$ by the chiral rotation, 
 with all the other field values set to zero. 
 
 The flat direction at tree level will be lifted by the quantum corrections, which 
generate a potential minimum for a non-zero value of $s^0$.  
As it was shown in Ref.~\cite{Gildener:1976ih}, 
with suitable renormalization conditions 
it is possible to take the value of the couplings 
such that $f_1=-\frac{f_2}{N_f}$ at a scale $\mu_{_{GW}}$.
Then the one loop potential $V_1$ can be expressed as 
\begin{align}
V_1(M) = \frac{1}{64\pi^2} \sum_{a=0}^{N_f^2-1} 
\left(
m_{s^a}^4(M) \left( \ln{\frac{m_{s^a}^2(M)}{\mu_{_{GW}}^2}} - \frac{3}{2} \right)
+
m_{p^a}^4(M) \left( \ln{\frac{m_{p^a}^2(M)}{\mu_{_{GW}}^2}} - \frac{3}{2} \right)
\right),
\end{align}  
where $m_{s^a}^2$ and $m_{p^a}^2$ are the mass functions 
for scalars and pseudoscalars:  
\begin{align}
m_{s^a}^2 = \frac{\partial^2 V_0(M)}{\partial (s^a)^2}, 
\quad \quad
m_{p^a}^2 =\frac{\partial^2 V_0(M)}{\partial (p^a)^2}. 
\end{align}  
By setting $p^a=0$ and $s^i = 0$, we define the $s^0$-dependent mass functions, 
which are given as 
\begin{align}
\label{eq:mass}
m_{s^0}^2(s^0) &= 0, 
& 
m_{s^i}^2(s^0) &= \left(f_1 +f_2\frac{3}{N_f}\right)(s^0)^2=\frac{2f_2}{N_f} (s^0)^2, 
\notag \\
m_{p^a}^2(s^0) &= 0, 
& 
m_{p^i}^2(s^0) &=0. 
\end{align}  
As mentioned above, 
if the GW condition is satisfied, 
the tree level potential is flat for the direction $s^0$, 
and the effective potential for $s^0$ is given 
\begin{align}
V_{\rm eff}(s^0) &=V_{\rm eff}(M)|_{M \to T^0
s^0}  = (V_0(M)+ V_1(M))|_{M \to T^0 s^0}
\notag \\
&= 
\frac{N_f^2-1}{64\pi^2} m_{s^i}^4(s^0)
\left( \ln{\frac{m_{s^i}^2(s^0)}{\mu_{_{GW}}^2}} - \frac{3}{2} \right), 
\label{eq:veff}
\end{align}  
The stationary condition for $s^0$ is 
obtained from the derivative of $V_{\rm eff}(s^0)$, 
\begin{align}
0&=\left.\frac{\partial V_{\rm eff}(s^0)}{\partial s^0}\right|_{s^0 \to \langle s^0 \rangle} 
\notag \\
&=
\frac{f_2^2}{4\pi^2}
\frac{N_f^2-1}{N_f^2}
\langle s^0\rangle^3
\left(
\ln{\frac{m_{s^i}^2(\langle s^0 \rangle)}{\mu_{_{GW}}^2}} - 1
\right).
\end{align}  
Thus a nonzero vev for $s^0$ is determined, 
\begin{align}
\label{eq:s0}
\ln{\frac{m_{s^i}^2(\langle s^0 \rangle)}{\mu_{_{GW}}^2}} = 1
\quad \Rightarrow \quad 
\langle s^0 \rangle = \sqrt{\frac{eN_f}{2f_2}} \mu_{_{GW}}.
\end{align}  
Due to no tree level contribution in the effective potential the stationary condition
is only determined from the loop correction, 
where 
the logarithmic term of a perturbative correction given by 
 ($\ln{\left(m_i^2(\langle M \rangle)/\mu_{_{GW}}^2\right)}-\frac{3}{2}$) 
does not depend on the inverse power of $f_2$ 
in contrast to a general CW potential, and stays small $(\mathcal{O}(1))$.
Therefore the perturbation analysis should work.

Thus we see that 
for the flat direction along the radial component $s^0$, 
the vacuum degeneracy can be lifted by the quantum correction, 
which give rise to the mass of $s^0$, 
which reads   
\begin{align}
\label{eq:ms02}
m_{s^0}^2 &= \frac{\partial^2 V_{\rm eff}(M)}{\partial (s^0)^2}\bigg|_{M=\langle M\rangle=T^0 \langle s^0\rangle} = \frac{f_2^2}{2\pi^2}
\frac{N_f^2-1}{N_f^2}  \langle s^0\rangle^2  
= \frac{e f_2}{4\pi^2} \left(
\frac{N_f^2-1}{N_f} \right) \mu_{_{GW}}^2. 
\end{align}  
Among the flavor non-singlet scalars $s^i$'s 
 the iso non-singlet scalars\footnote{
As for the iso-singlet scalars, 
the mass function is not positive definite for $N\geq 3$ at one loop level. 
This problem can be avoided by adding a small explicit breaking mass term, 
and we will discuss phenomenological implication of such effect.} 
have a common 
positive definite mass, 
which reads from Eq.(\ref{eq:mass}) and Eq.(\ref{eq:s0}):
\begin{align}
\label{simass}
m_{s^i}^2 &= \left. \frac{\partial^2 V_{\rm eff}(M) 
}{\partial (s^i)^2}\right|_{M=\langle M \rangle}  = 
m_{s^i}^2(\langle s^0 
\rangle)
= e \mu_{_{GW}}^2.
\end{align}  
From Eq. (\ref{simass}) and Eq.(\ref{eq:ms02}) we have
\begin{align}
\frac{m_{s^0}^2}{ m_{s^i}^2}= \frac{f_2 N_f}{4\pi^2} \left(
\frac{N_f^2-1}{N_f^2} \right)\quad \left(\ll 1\quad {\rm for}\,\,  \frac{f_2 N_f}{4\pi^2} \ll 1 \right)\,,
 \end{align}
where the 't Hooft coupling $f_2 N_f$ is fixed in the anti-Veneziano limit $N_c \rightarrow \infty$ with $N_f/N_c=$ fixed $(>1)$ of the underlying large $N_f$ QCD. 

As for the $N_f^2$ pseudoscalars, $p^a$'s, 
they still remain massless due to 
the other flat directions, which correspond to the pion fields ($\pi^a$), 
the massless NG 
bosons of the chiral symmetry breaking $U(N_f)_L\times U(N_f)_R/U(N_f)_V$.

These massless NG bosons can have mass as follows: 
In the underlying large $N_f$ QCD, the axial $U(1)_A$ anomaly for the $N_f$ fermions 
should be responsible for the mass of flavor-singlet ''$\eta'$'' , 
which could be  even much heavier than in the ordinary QCD  particularly in the anti-Veneziano limit $N_c \rightarrow \infty$
with $N_f/N_c \gg 1$ (the limit corresponding to the walking technicolor) \cite{Matsuzaki:2015sya}, 
in good agreement with the preliminary lattice results \cite{Aoki:2016fxd}.
Under this circumstance we simply assume this $\eta^\prime$ as super heavy to be decoupled in the linear sigma model treatment for $N_f=8$ in the present paper.
\footnote{  
The effect of $U(1)_A$ breaking effect in the linear sigma model at tree-level
has been recently discussed~\cite{Meurice:2017zng} to understand the lightness of the singlet scalar on the lattice \cite{Aoki:2016fxd}.
}
The rest of $N_f^2-1$ NG bosons in the  $N_f=8$ one-family technicolor model are no longer massless; The chiral symmetry $SU(8)_L\times SU(8)_R$ are explicitly broken down to $SU(2)_L\times U(1)_Y$ through the SM gauge interactions (QCD and electroweak) and the ETC 
gauge interactions, so that 60 NG bosons acquire the mass (becoming pseudo NG bosons), leaving us with only three of them which are of course eaten into $W/Z$ bosons via the standard Higgs mechanism~\cite{Farhi:1980xs}. Masses of all these 60 pseudo NG bosons are enhanced in the walking theory by the large anomalous dimension $\gamma_m \simeq 1$ (see \cite{Matsuzaki:2015sya} and references therein). 
 In this paper,  instead of explicitly formulating these effects, we shall employ a handy way by simply introducing ad hoc mass breaking at tree level into the linear sigma model in subsection \ref{sec:mass}.
Although it is very simple to illustrate the essential feature of the phase transition and the gravitational waves, 
it has some phenomenological problem. 
For completeness we then demonstrate a more realistic soft breaking term in the Farhi-Susskind one-family model with $N_f=8$, which are free from such a problem.

\subsection{Equivalence to the dilaton potential in the walking technicolor}\label{sec:dxpt}

We discuss the possibility that this light scalar $s^0$ in the above frame may be
regarded as the pseudodilaton, the pseudo NG 
boson 
associated with the approximate scale invariance, which is nothing but a technidilaton in the context of  the walking technicolor.

In fact we notice that the scalar potential in Eq.(\ref{eq:veff}) is equivalent to the CW-like
dilaton 
potential for the walking technicolor, Eq.~(3) (with the spurion set to be $S=1$) in Ref.~\cite{Matsuzaki:2013eva}, 
\begin{align}
\label{eq:dilaton}
\mathcal{L}
_{(2){\rm hard}} = 
-\frac{m_\phi^2 F_\phi^2}{4} 
\chi^4  \left(
\ln \chi 
-\frac{1}{4} 
\right) =-\frac{m_\phi^2}{4 F_\phi^2} 
\sigma^4  \left(
\ln\frac{\sigma }{F_\phi}
-\frac{1}{4} 
\right), \quad  \left(F_\phi \equiv \langle \sigma \rangle \right)
\end{align}  
where  
$ \sigma (x) \equiv 
F_\phi\cdot  \chi(x)$  with $\chi(x)\equiv  e^{\phi(x)/F_\phi}$ is a chiral singlet field 
with the canonical dimension 1 under the scale transformation and is
related to the dilaton field $\phi(x)$ s.t. $\langle \phi(x)\rangle=0$ ($\langle \chi(x)\rangle=1)$),\footnote{ The scale (dilatation) transformations for these fields are 
$ \delta_D \sigma =(1 +x^\mu \partial_\mu) \sigma$, 
$\delta_D \chi=(1+x^\mu \partial_\mu) \chi$, 
$\delta_D \phi= F_\phi +x^\mu \partial_\mu\phi$. 
Although $\chi$ is a dimensionless field,
it transforms as that of dimension 1, while $\phi$ having dimension 1 transforms as the dimension 0, instead.
}
 with the dilaton decay constant, 
$F_\phi (=\langle \sigma \rangle)$, 
and $m_\phi (\simeq 125$ GeV) is 
the mass of the (pseudo) dilaton $\phi$ to be identified with the 125 GeV Higgs. 
 It indeed  reproduces the PCDC relation 
 \begin{eqnarray}
 m_\phi^2 F_\phi^2 =-F_\phi\langle 0| \partial^\mu D_\mu |\phi\rangle =- d_\theta \langle \theta_\mu^\mu\rangle=- d_\theta \langle \delta_D \mathcal{L}^S_{(2){\rm hard}} \rangle\,, 
 \label{PCDC}
 \end{eqnarray}
with $d_\theta=4$.
The trace anomaly $\theta^\mu_\mu \ne 0$ measuring the explicit breaking of the scale symmetry comes from 
the spontaneous breaking of the chiral symmetry and the scale symmetry  in the scale-invariant dynamics (ladder SD equation, $\grave{a}$ la anti-Veneziano limit) for the underlying large $N_f$ QCD, which generates the spontaneous breaking mass scale actually provided by 
the cutoff regulator or the renormalization point. 
See 
 \cite{Matsuzaki:2015sya} and references therein. 
\footnote{The same effective potential was directly derived in the  
the ``quenched ${\rm QED}_4$''~\cite{Miransky:1996pd} up to  trivial factors of $N_c$ and $N_f$, 
resulting in the same PCDC relation. 
}
 
Now for the present  case of the linear sigma model we may write $\sigma^2=2  {\rm Tr} (M^\dagger M)$, with $\sigma$ being the field in the ray (radial) direction in the tree potential as in the GW arguments and actually the same field as $\sigma$ in Eq.(\ref{eq:dilaton})  to be obviously chiral singlet having the canonical scale dimension 1.
The trace expression  can always be rotated into $ 2  {\rm Tr} (M^\dagger M)\rightarrow (s^0)^2$ by the chiral transformation into a particular frame $s^i, p^a \rightarrow 0$, so that $\sigma^2=(s^0)^2$, namely 
we can identify $\sigma$ with $s^0$. 
Thus the two free parameters $f_2$ and $\langle s^0\rangle$ (or, $\mu_{_{GW}}$) in our model 
are related to the dilaton mass $m_{s^0}=m_\phi \simeq 125$ GeV and its decay constant $F_\phi=\langle s^0\rangle$ in the dilaton potential.
 The value of $F_\phi$ of a technidilaton as the 125 GeV Higgs, $m_\phi=125$ GeV, is estimated \cite{Matsuzaki:2015sya} through the best fit to the LHC data for the 125 GeV Higgs as
 $F_\phi\simeq 1.25 {\rm TeV} \simeq 5\, v_{_{\rm EW}}$ 
 ($N_f=8, N_c=4$), which is also consistent with the ladder estimate for the trace anomaly  $\langle \theta^\mu_\mu\rangle= 4 \langle\theta^0_0\rangle$ and $v_{\rm EW}$ in the large $N_f$ QCD. 

Comparing Eq.(\ref{eq:dilaton}) with Eq.(\ref{eq:veff}), 
we find the following identifications for $N_f=8$ in the frame mentioned above:
\begin{align}
\label{benchmark}
s^0&= 
\sigma = F_\phi\cdot  e^{\phi/F_\phi}
, \ \ \ 
\langle s^0 \rangle 
= \langle \sigma \rangle=F_\phi \simeq 5\cdot v_{_{\rm EW}} \simeq 1.25\, {\rm TeV}, 
\notag \\
f_2^2 &= 
 2\pi^2 \frac{m_\phi^2}{F_\phi^2} \frac{N_f^2}{N_f^2-1} \simeq \left(0.45\right)^2 
 \, \ll 1,
\end{align}  
where the numerical values for $F_\phi$ and $f_2$ are for the best fit value to the LHC data for the 125 GeV Higgs 
mentioned above, and will be our bench mark value in the later discussions. 
Thus $s^0$ (more precisely the frame-independent radial field $\sigma$, the flavor singlet scalar) plays a role of 
the 
pseudodilaton $\phi=\ln (\sigma/\langle \sigma \rangle)$ 
 as originally referred to as  ``scalon'' in Ref.~\cite{Gildener:1976ih}.
Note that $F_\phi^2\sim N_f N_c \sim N_f^2$ in the anti-Veneziano limit of the underlying large $N_f$
 QCD and hence $f_2 N_f=$ fixed in that limit. 
\footnote{
For our bench mark value in Eq.(\ref{benchmark}), we would have $m_{s^0}^2/m_{s^i}^2\simeq (0.3)^2$. Within the linear sigma model, this hierarchy  can be further
enlarged by introducing a large tree-level  explicit mass term  for $p^a$ mocking up  the explicit breaking due to the SM and ETC gauge interactions in the walking technicolor.  See the next subsection.
}

Thus the same symmetry relation should hold in our potential. 
Our result indicates that the low energy effective theory of large $N_f$ QCD 
can be approximately described 
by a linear sigma model having the scale invariance in  the weak coupling regime $f_2 \simeq 0.45 (<1)$ and $|f_1|=|-f_2/N_f|\simeq 0.056\ll 1$, where
the perturbative picture works at the composite level.

For discussing the electroweak phase transition,
one might literally gauge the linear sigma model kinetic term in Eq.(\ref{linearmodel}), with the electroweak $W/Z$ bosons, 
in which case the $W/Z$ boson mass scale would be given by $\langle s^0 \rangle=v_{_{\rm EW}}=246\, {\rm GeV}$ 
in apparent contradiction with our bench mark value 
$\langle s^0 \rangle=F_\phi \simeq 5 \, v_{_{\rm EW}}$ in Eq.(\ref{benchmark}).
However, the contradiction is avoided by noting
that Eq.(\ref{benchmark}) is based on the dilaton effective theory in Ref.~\cite{Matsuzaki:2013eva}, 
whose kinetic term is
\begin{eqnarray}
\label{NLS}
{\cal L}^{(\rm inv)}_{(2)}=\frac{F_\phi^2}{2} (\partial_\mu \chi)^2 + \frac{F_\pi^2}{4} \chi^2 \, {\rm Tr} \left[\partial_\mu U \partial^\mu U^\dagger \right]\,, \quad U(x) \equiv
 \exp \left(\frac{2 i \pi(x)}{F_\pi}\right)\,, \quad \pi(x)\equiv \sum_{i=1}^{N_f^2-1} \pi^i(x) T^i\,.
\end{eqnarray}
where $F_\pi$ is the decay constant of NG boson $\pi$ to be eaten into $W/Z$
($U(1)_A$ is explicitly broken by the axial anomaly in the underlying theory and is  irrelevant to the $W/Z$ mass anyway).
This has the scale symmetry and chiral $SU(N_f)_L\times SU(N_f)_R$, both spontaneously broken,
the same symmetry structure as that in the underlying large $N_f$ QCD,
where $F_\phi$ is in general different from $v_{_{\rm EW}}= \sqrt{N_f/2} \cdot F_\pi$ as shown in the ladder calculations $\grave{a}$
la anti-Veneziano limit.
Indeed the kinetic term of $\pi(x)$ in Eq.(\ref{NLS}) is normalized to be canonical for any value of $F_\pi$.
\footnote{
The linear sigma model kinetic term Eq.(\protect\ref{linearmodel}) corresponds to only a particular choice 
$F_\pi= F_\phi/\sqrt{N_f/2}$, where Eq.(\protect\ref{NLS}) rewritten in terms of
$\tilde M=\sigma T^0 \cdot U=F_\phi \chi T^0\cdot U$ takes precisely the same form as  Eq.(\protect\ref{linearmodel}) for $M$.
Alternatively, any complex matrix $M$ can be written in the polar decomposition $M=H\cdot U$,
where $H= \sum_{a=0}^{N_f^2-1} \tilde s^a T^a$ is the Hermitian matrix
and $U$ the unitary to be identified with the $U$ in Eq.(\protect\ref{NLS}), then 
$
{\rm Tr} [\partial_\mu M^\dagger \partial^\mu M]=
{\rm Tr} [(\partial_\mu H)^2] + {\rm Tr} [H^2 \partial_\mu U \partial^\mu U^\dagger]
$.
Although $H$ contains massive $N_f^2-1$ flavor non-singlet scalar $\{s^i\}\sim  \{\tilde s^i\}$
in addition to the light singlet scalar (pseudodilaton) $s^0\sim \tilde s^0$, these massive components
may be ignored in the decoupling limit 
as to set  $H\sim s^0 T^0 =F_\phi \chi T^0$ at one loop (at higher loops in that limit,
counter terms of the high dimensional operators will destroy such a simplest result,
leading to the generic case Eq.(\protect\ref{NLS})). 
}
Actually, after the spontaneous symmetry breaking  due to the log potential arising from quantum corrections, the linear sigma model 
kinetic term Eq.(\ref{linearmodel}) may no longer be valid generically, particularly for the NG boson parts irrelevant to the potential, which is given by the second term in Eq.(\ref{NLS}) as a most general  effective theory of the underlying theory. 
Based on Eq.~(\ref{NLS}), the dilaton condensate $\langle s^0\rangle$ at zero temperature vacuum
is not identified with the electroweak scale $v_{_{\rm EW}} = \sqrt{N_f/2} \cdot F_\pi = 246$ GeV
and taken to be much higher dilaton scale ($F_{\phi}\sim \mathcal{O}(1)$ TeV)
favored in the walking technicolor model~\cite{Matsuzaki:2015sya}.
By using $F_{\phi}$, the top quark mass squared is expressed as
\begin{align}
m^2_t = \frac{y_t}{2}v^2_{_{\rm EW}}
= \frac{y_t}{2}\frac{v^2_{_{\rm EW}}}{m^2_{\phi}}\frac{m^2_{\phi}}{F^2_{\phi}}\langle s^0\rangle^2
\ ,\label{eq:mmtop}
\end{align}
where $y_t\sim \mathcal{O}(1)$ denotes the top Yukawa coupling.
In comparison, the squared mass of the flavor non-singlet scalars,
which are the degrees of freedom contributing to the effective potential, reads,
\begin{align}
m^2_{s^i} = \frac{2\sqrt{2}\pi}{N_f}\sqrt{\frac{N_f^2}{N_f^2 - 1}}\frac{m_{\phi}}{F_{\phi}}\langle s^0\rangle^2
\ ,\label{eq:mms_w_fphi}
\end{align}
using Eqs.~(\ref{eq:mass}) and (\ref{benchmark}).
For $N_f = 8$, $m^2_{s^i}\sim \mathcal{O}(1)\times (m_{\phi}/F_{\phi})\langle s^0\rangle^2$.
Comparing this with Eq.~(\ref{eq:mmtop}), the top quark mass squared is found to be
much smaller: $m^2_t / m^2_{s^i}\sim \mathcal{O}(m_{\phi} / F_{\phi})\ll 1$.
The same order counting holds in the squared masses of $W$ and $Z$ bosons compared with $m^2_{s^i}$.
Note that the effective potential at zero temperature is proportional to the fourth power of masses ($m^4_{_X}$, $X = t$ or $s^i$),
and the absolute value of the thermal function $J_{B/F}$ ({\em c.f.} Eqs.~(\ref{eq:JB}) and (\ref{eq:JF}))
is an increasing function of its argument $m^2_{_X} / T^2$.
Therefore, the SM contribution to the effective potential is subdominant.
Moreover, the degrees of freedom of top quarks (= 12) and weak bosons (= 6 + 3)
are smaller than those of the flavor non-singlet scalars $N_f^2 - 1 = 63$.
This results in further suppression of the top quark and weak boson effects.
In Sec.~\ref{sec:discuss_top},
we will numerically confirm that the top quarks do not modify the qualitative feature of gravitational waves in our model.

\subsection{Soft chiral and scale symmetry breaking term} \label{sec:mass}

As mentioned earlier, the large $N_f$ QCD in the walking regime would produce many massless NG bosons for the spontaneously broken large chiral symmetry
$U(N_f)_L\times U(N_f)_R$, so does our linear sigma model with massless $N_f^2$ pseudoscalar $p^a$.
In the underlying large $N_f$ QCD, the flavor-singlet one ``$\eta^\prime$'' should acquire mass form the axial anomaly. When the theory is applied to the actual 
walking technicolor coupled to 
the SM gauge and ETC gauge interactions, these gauge interactions break explicitly the large chiral symmetry to give mass to other pseudoscalar NG  bosons. In the case at hand $N_f=8$, 
while the three among the rest $63$ NG bosons  should be eliminated by the Higgs mechanism of the 
gauged $SU(2)_L\times U(1)_Y$ symmetry, 
all the remaining 60 NG bosons should obtain 
mass from the SM and the ETC gauge interaction similarly to the $\pi^+-\pi^0$ mass difference in the usual QCD~\cite{Farhi:1980xs}.  Although these couplings are weak to be perturbative, the large anomalous dimension $\gamma_m \simeq 1$ of the walking technicolor
enhances all these masses to the TeV scale. 
For details see Ref. \cite{Matsuzaki:2015sya} and the references cited therein.

In the following analysis, instead of discussing these effects explicitly, 
 here we 
consider an extension of the potential by 
simply adding some tree level mass term in the potential mocking up 
these explicit breakings. 
(For a study of the SM contributions to mass terms 
in the (non-conformal) linear sigma model,  see Ref.~\cite{Kikukawa:2007zk}).
Among various possibilities of the chiral symmetry breaking mass terms, 
we consider 
the following 
mass term
\footnote{ 
While the above-mentioned SM and ETC gauge interactions leave the three NG bosons  be exact massless to be eaten into $W/Z$ bosons, this mass term gives mass also  to  such  NG bosons. However, for the purpose of the present paper to see the phase transition nature of the scalar sector relatively independent of  $W/Z$ mass generation, this would not   be a serious problem.
For completeness we shall later present a  refined treatment on Eq.(\ref{soft})  in a concrete case of $N_f=8$ in the Farhi-Susskind model,  which yields the same explicit breaking  as that due to the SM and ETC gauging to keep massless the three NG bosons to be eaten into $W/Z$ bosons, while giving mass to all other 61 NG bosons without affecting the Higgs mass.
\label{breakingterm}
}
\begin{align}
\label{soft}
V_{\rm soft
} &
= \frac{(\Delta m_p)^2}{2} \sum_{a=0}^{N_f^2-1}(p^a)^2. 
\end{align}  
Then the tree-level potential is given by $V(M)=V_0(M)+V_{\rm soft
}(M)$.  
Compared 
with Eq.~(\ref{eq:mass}), 
the $s^0$-dependent mass functions for this model at tree level modify only $p^a$ part 
\begin{align}  
\label{eq:massmp}
 m_{p^a}^2(s^0, \Delta m_p
) 
= (\Delta m_p)^2, 
\end{align} 
while not $s^0$ and $s^i$:
\begin{align}  
\label{eq:massmp2}
m_{s^0}^2(s^0, \Delta m_p
) &= 0, \quad 
 m_{s^i}^2(s^0, \Delta m_p
) = 
\frac{2f_2}{N_f} (s^0)^2.
\end{align} 

Note that this explicit breaking $p^a$ mass term in Eq.(\ref{soft}) is completely different from the conventional one in the linear sigma model (mocking the 
quark mass breaking), 
\begin{align}
V_m =-c \sigma=-c s^0,  \quad \left(c=(\Delta m_p)^2 \langle s^0\rangle\right),
\label{linearbreaking}
\end{align}
which yields via stationary condition the same term as Eq.(\ref{soft}) but also affects the mass of $s^0$ already at tree level in sharp contrast to Eqs.(\ref{eq:massmp}) and (\ref{eq:massmp2}). Eq.(\ref{soft}) thus resembles the explicit chiral symmetry breakings by the SM gauge and ETC gauge interactions which give mass to the pseudo scalar NG bosons (technipions) while not affecting the mass of the technidilaton as the Higgs $s^0=\sigma$.

At one loop level, we immediately see that this explicit breaking Eq.(\ref{soft}) does not change the stationary condition and the mass for $s^0$ 
from the original potential,
since there is no $s^0$ dependence in the mass functions for the pseudoscalars (no $(s^0)^2 (p^{(a)})^2$ couplings
under the GW condition Eq.(\ref{GWcond})) 
and the mass functions for the scalars are the same even at one loop.\footnote{
One might suspect that  Eq.(\ref{soft})  would change in principle the mass of Higgs as the pseudodilaton $s^0=\sigma$, since it breaks the scale symmetry explicitly.
Although it in fact gives an additional contribution to the trace anomaly  $\theta_\mu^\mu$, the PCDC relation
 relevant to $\langle \theta_\mu^\mu\rangle$
, Eq.(\ref{PCDC}),
 is actually unchanged and so is the mass of 
$s^0=\sigma$ as the Higgs, 
since 
the additional contributions to the relation are $\langle \Delta \theta_\mu^\mu\rangle \sim \langle (p^{(a)})^2\rangle =0$. This is indeed consistent with the  explicit one loop calculations given here.}
Hence even at  one loop the mass of $s^0$  is not affected by the explicit breaking Eq.(\ref{soft}) and remains the same as Eq.(\ref{eq:ms02}):
\begin{align}
m_{s^0}^2 &= \frac{f_2^2}{2\pi^2}
\frac{N_f^2-1}{N_f^2}  \langle s^0\rangle^2  
= \frac{e f_2}{4\pi^2} \left(
\frac{N_f^2-1}{N_f} \right) \mu_{_{GW}}^2.
\end{align}
Thus the mass of the Higgs $s^0=\sigma$ is not affected by this explicit breaking Eq.(\ref{soft}) at quantum level, similarly to the explicit breaking by the SM and ETC gauging in the $N_f=8$  technicolor model.

On the other hand,
this potential 
gives additional mass to the 
non-singlet scalars through the quantum effect.
Then the other mass spectra are modified as follows 
\begin{align}
m_{s^i}^2 =& 
e\mu_{_{GW}}^2 
+ \frac{f_2}{32\pi^2} 
\biggl\{ 
2N_f (\Delta m_p)^2 \left( \ln{\left(\frac{(\Delta m_p)^2}{\mu_{_{GW}}^2}\right)} -1 \right)
+
e f_2 \mu_{_{GW}}^2 \left(9+ \ln{\left(\frac{(\Delta m_p)^2}{\mu_{_{GW}}^2}\right)} \right)
\sum_{j=1}^{N_f^2-1} (d_{jji})^2 
\biggl\} 
\notag 
\\
 m_{p^0}^2 =& (\Delta m_p)^2 
+ \frac{4f_2}{32\pi^2} 
\left(\frac{N_f^2-1}{N_f}\right) (\Delta m_p)^2  
\left( \ln{\left(\frac{(\Delta m_p)^2}{\mu_{_{GW}}^2}\right)} -1 \right)
\notag \\
m_{p^i}^2 =& (\Delta m_p)^2 
+ \frac{f_2}{32\pi^2} 
\left(2N_f-\frac{4}{N_f}\right)  (\Delta m_p)^2\left( \ln{\left(\frac{(\Delta m_p)^2}{\mu_{_{GW}}^2}\right)} -1 \right), 
\label{oneloopNGmass}
\end{align}  
where $d_{ijk}$ is the totally symmetric tensor  
\begin{align}  
d_{ijk} = 2{\rm Tr} [\{ T^i,T^j\} T^k],
\end{align}  
for the generators of $SU(N)$.
As shown here, 
we see that the iso-singlet scalar mass can be negative
in the vicinity of the zero mass $\Delta m_p \to 0$, 
since only the iso-singlet sector has  
a non-zero value for 
$d_{jji}$, a log term which is proportional to $d_{iij}$ 
becomes infinite in the limit $\Delta m_p\to 0$.
Thus the mass function for the iso-singlet (not flavor-singlet) scalar 
becomes well defined if $\Delta m_p> \mu_{_{GW}}$ for 
$N_f>2$.

\begin{figure}[!tbp]
\begin{center}
\includegraphics[width=3in]{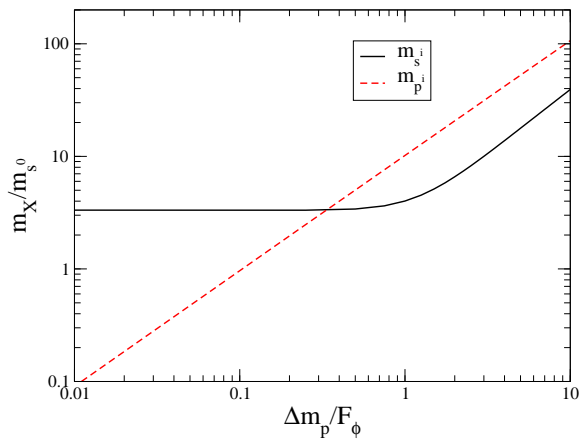}
\end{center}
\caption{$\Delta m_p$ dependent mass ratio for $N_f=8$ with $f_2 = 0.45$.}\label{fig:ratiop}
\end{figure}

The resulting mass ratios obtained from the explicit mass terms 
are shown in Fig.~\ref{fig:ratiop}
As shown here, 
a large mass hierarchy between the flavor singlet and non-singlet scalar masses 
is obtained without a tree-level scalar mass. 
While both the $m_{p^i}$ and $m_{s^i}$ have a linear dependence on
the mass $(\Delta m_p)^2$, the coefficient for  $m_{s^i}$ is suppressed 
by a one-loop factor compared to $\Delta m_p$. 
Following again \cite{Matsuzaki:2013eva}, 
let us assume that $s^0\sim \sigma$ can be regarded as the pseudo dilaton, 
which corresponds to the Higgs boson with mass $125$ GeV in SM, 
and the vev $\langle s^0 \rangle =\langle \sigma \rangle $ corresponds to the dilaton decay constant $F_\phi$. 
As increasing $\Delta m_p
$ we see a large mass 
hierarchy between light pseudo dilaton and the others. 
Assuming $\Delta m_p 
 \sim F_\phi \simeq 1.25 $ TeV 
as a typical new physics scale beyond SM, 
we can obtain a mass of $\mathcal{O}(1)$ TeV for technihadrons.

Thus the simple-minded explicit breaking Eq.(\ref{soft}) mocks up the 
explicit chiral symmetry breaking due to the SM and ETC gauging in the ETC model, in the sense that it gives mass to
the NG bosons $p^a$'s 
without changing the Higgs mass $\sigma=s^0$. Although it is enough for the present paper to study of the phase transition relevant to the gravitational waves, 
however, as noted before (see footnote \ref{breakingterm}) the resultant NG boson mass, Eq(\ref{eq:massmp}) and (\ref{oneloopNGmass}), 
also includes non-zero mass of the  3 NG bosons 
required massless so as to be eaten into $W/Z$ bosons. This as it stands is  a phenomenological disaster. 

For the sake of the reader's concern on this point, we now discuss
a more realistic model of the walking technicolor,  
e.g., the Farhi-Susskind one-family model with $N_f=8$~\cite{Farhi:1980xs},
in  which we can explicitly construct a (though somewhat complicated) desired
breaking term without such a problem as in Eq.(\ref{soft}).
In this case the chiral symmetry is 
$SU(2N_D)_L \times SU(2N_D)_R$, 
where $N_D=N_f/2=4$, the number of the weak doublets 
consisting of three techniquark $Q_c$ $(c=1,2,3)$ 
and one technilepton doublet $L$. 
It is convenient to use a different basis of the following generators $X^a$ $(a=0, \cdots, 63)$ 
of $U(8)$ Lie algebra~\cite{Farhi:1980xs, Jia:2012kd},
\begin{align}  
X^0 &=
\frac14 1_{8 \times 8}, 
\nonumber \\ 
X^1
&=
X_P
= 
\frac{1}{4 \sqrt{3}}
\left( 
\begin{array}{c|c} 
1_{6 \times 6}  & 0 \\ 
\hline 
0 & -3\cdot 1_{2 \times 2} 
\end{array}
\right) \,, 
\qquad &
X^{a=2, 3, 4}
&=
X^i_P = 
\frac1{2 \sqrt3}
\left( 
\begin{array}{c|c} 
\tau^i \otimes 1_{3\times 3}  &  0 \\ 
\hline 
0 & -3 \cdot \tau^i 
\end{array}
\right)  \,, 
\nonumber \\ 
X^{a=5, \cdots, 12}
&=
X_{\theta_i} 
= 
\frac{1}{2 \sqrt{2}}
\left( 
\begin{array}{c|c} 
1_{2\times 2} \otimes \lambda_i  & 0 \\ 
\hline 
0 & 0  
\end{array}
\right)  \,, 
\qquad &
X^{a=13, \cdots, 36}
&=
X_{\theta_j}^i 
= 
\frac1{\sqrt2}  
\left( 
\begin{array}{c|c} 
\tau^i \otimes \lambda_j  & 0 \\ 
\hline 
0 & 0 
\end{array}
\right)  \,, 
\nonumber \\ 
X^{a=37, 38, 39}
&=
X^{(1)}_{T c}  
=  
\frac14
\left( 
\begin{array}{c|c} 
0 & 1_{2 \times 2} \otimes \xi_c    \\ 
\hline 
1_{2 \times 2} \otimes \xi_c^\dag  &  0
\end{array}
\right) 
\,, \qquad &
X^{a=40, 41, 42}
&=
X^{(2)}_{T c} 
= 
\frac14
\left( 
\begin{array}{c|c} 
0 & - i \cdot 1_{2 \times 2} \otimes \xi_c    \\ 
\hline 
i \cdot 1_{2 \times 2} \otimes \xi_c^\dag  &  0
\end{array}
\right) 
\,, 
\nonumber \\ 
X^{a=43, \cdots, 51}
&=
X^{(1)i}_{T c} 
=  
\frac12
\left( 
\begin{array}{c|c} 
0 & \tau^i \otimes \xi_c    \\ 
\hline 
\tau^i \otimes \xi_c^\dag  & 0 
\end{array}
\right) 
\,, \qquad &
X^{a=52, \cdots, 60}
&=
X^{(2)i}_{T c}  
= 
\frac{1}{2}
\left( 
\begin{array}{c|c} 
 & - i \tau^i \otimes \xi_c    \\ 
  \hline 
i \tau^i \otimes \xi_c^\dag  &  
\end{array}
\right) 
\,, 
\nonumber \\ 
X^{a=61, \cdots, 63}
&=
X_{\rm eaten}^i 
=
\frac12 \left( 
\begin{array}{c|c} 
\tau^i \otimes 1_{3\times 3}  & 0 \\ 
\hline 
0 & \tau^i 
\end{array}
\right)  \,, 
\end{align}  
where $\xi_c$ is the three-dimensional unit vector for $SU(3)_{\rm color}$, 
and $\tau^i$ and $\lambda_i$ are the Pauli and Gell-Mann matrices, respectively.
The generators have the same normalization of $T^a$ as 
\begin{align}
{\rm Tr} [X^a,X^b] = 
\delta^{ab}/2.
\end{align}  
The technipions $\pi^a$ constructed from $p^a$ 
can be classified by the 
weak iso-spin and the color charges as follows,  
\begin{align}  
\sum_{a=0}^{63} \pi^a(x) X^a 
=&
p^0 X^0 +  P X_P 
+ \sum_{i=1}^3 P^i X^i_{P}
+ \sum_{j=1}^8 \theta_j X_{\theta j}  
+ \sum_{i=1}^3 \sum_{j=1}^8 \theta^i_j X_{\theta j}^i 
\nonumber \\  
& 
+\sum_{c=1}^3 \sum_{i=1}^3 \left[ T_c^{(1)i} X_{T c}^{(1)i} + T_c^{(2)i} X_{T c}^{(2)i} \right] 
+
\sum_{c=1}^3 \left[ T_c^{(1)}  X_{T c}^{(1)}  + T_c^{(2)} X_{T c}^{(2)} \right] 
+\sum_{i=1}^3 \pi_{\rm eaten}^i X_{\rm eaten}^i,
\end{align}  
where $\pi_{\rm eaten}^i(x)$ are the unphysical NG bosons, 
which should be eliminated by the Higgs mechanism of the gauged 
$SU(2)_L \times U(1)_Y$ symmetry.
It is known that in the one-family model the remaining 61 NG bosons 
(60 NG bosons plus one singlet pseudoscalar) 
should be massive due to the 
the SM and ETC interactions which break the full chiral symmetry. 
We then consider modeling of the pseudo NG boson masses
by adding chiral symmetry breaking mass terms,
\begin{align}
\label{soft2}
V_{\rm soft} 
=&
 \frac{(\Delta m_{p^0})^2}{2} (p^0)^2
+\frac{(\Delta m_P)^2}{2} (P)^2
+\frac{(\Delta m_{\theta_i})^2}{2} \sum_{i=1}^8 (\theta_i)^2
+\frac{(\Delta m_{\theta_i})^2}{2} \sum_{i=1}^3 \sum_{j=1}^8 (\theta_j^i)^2
+\frac{(\Delta m_{\theta_i})^2}{2} \sum_{i=1}^3 \sum_{j=1}^8 (\theta_j^i)^2
\nonumber \\  
& 
+(\Delta m_{T_c})^2 \sum_{c=1}^3 \left|T_c\right|^2
+(\Delta m_{T_c^i})^2 \sum_{c=1}^3 \sum_{i=1}^3 \left|T_c^i\right|^2,
\end{align}  
where $T_c^i$ and $T_c$ are the color-triplet technipions, 
\begin{align}  
T_c = \frac{T_c^{(1)}-iT_c^{(2)}}{\sqrt{2}},
\qquad
T_c^i = \frac{T_c^{(1)i}-iT_c^{(2)i}}{\sqrt{2}}. 
\end{align}  
Again we note that the soft mass terms should have 
a different origin from the effect of the current technifermion masses
which give rise to a conventional linear term of 
in the potential Eq.(\ref{linearbreaking}).
Comparing Eq.(\ref{soft2}) with 
Eq.~(\ref{soft}), 
we have non-degenerate mass parameters for the technipion. 
In the typical walking technicolor with $\gamma_m \simeq 1$, the scale of the masses for 
$61$ NG bosons are of 
order of 
TeV, see \cite{Matsuzaki:2015sya} and references therein for details. 
Having such soft mass parameters  
we may have a rich hadron structure for the masses of the 
non-singlet scalars ($s^{i=1,\cdots, 63}$) 
and the pseudoscalars ($p^{i=0,\cdots,60}$).  
The remaining NG bosons $\pi_{\rm eaten}^i$ stay massless at tree-level,
to be eaten into $W/Z$ bosons as desired.

As in the case of Eq.~(\ref{soft}), 
owing to the quadratic form of the soft breaking terms, 
the $s^0$-dependent mass functions for the scalars do not 
depend on the tree-level technipion mass parameters, 
and the mass functions $m_{p^a}^2$ do not depend on $s^0$. 
Therefore in our model
the additional potential term $V_{\rm soft}$ 
does not change the dilaton potential given in Eq.~(\ref{eq:dilaton})
at the one loop level, 
and the dilatonic nature of the flavor singlet scalar:
\begin{align}  
\left.\frac{\partial V_{\rm eff}}{\partial s^0 }\right|_{s^0=0} = 0, 
\quad \quad 
\left.\frac{\partial^2 V_{\rm eff}}{\partial (s^0)^2 }\right|_{s^0=0} =0, 
\end{align}  
remains intact.

From the result, 
we can expect that the first-order phase transition at finite temperature 
can naturally occur
since a positive quadratic curvature at the origin 
can be easily generated 
through the thermal effect.
We will examine the thermal phase transition for the dilaton potential in the next section.
We also note that the perturbative correction of the logarithmic term is small in the 
 effective potential for the radial direction $V_{\rm eff}(s^0\sim \sigma)$  
\begin{align}
\label{eq:veffm0}
V_{\rm eff}=\frac{N_f^2-1}{64\pi^2} 
m_{s^i}^4(s^0) 
\left( \ln{ \frac{m_{s^i}^2(s^0)}{\mu_{_{GW}}^2}} -\frac{3}{2}
\right) + C,
\end{align}
where $C$ is a constant, 
 so that the GW mechanism still works even 
if we have an additional 
parameter 
$(\Delta m_p)^2$.

\section{Finite temperature potential and electroweak phase transition}

In this section we investigate the effective potential 
at finite temperature $T$~\cite{Weinberg:1974hy, Dolan:1973qd}
and study the thermal phase transition of the chiral and scale symmetry.
To evaluate the finite temperature effective potential, 
we use the one loop contribution of the thermal effect $V_{1,T}(s^0, T)$
\begin{align}  
\label{eq:V1T}
V_{1,T}(s^0,T) = \frac{T^4}{2\pi^2}
\sum_{a=0}^{N_f^2-1} \left\{ J_B\left(m_{s^a}^2(s^0, \Delta m_p)/T^2\right) 
+ J_B\left(m_{p^a}^2(s^0, \Delta m_p)/T^2\right) \right\} ,
\end{align}  
where $J_B$ is the bosonic thermal function 
\begin{align}
\label{eq:JB}
J_B(x)= \int_0^\infty t^2 
\ln{\left(1-e^{-\sqrt{t^2+x}}\right)} dt\ .
\end{align}  
We directly evaluate Eq.~(\ref{eq:JB}) without expanding in its argument $m_{s^a}^2/T^2$.
It is known that usual perturbative expansion valid at zero temperature will break down at high temperature
where the higher loop contributions, in particular quadratic divergent loops (daisy diagrams), can not be neglected.
We include the daisy diagrams which correspond to replace the mass functions
in both zero and finite temperature effective potential ($V_1$ and $V_{1,T}$)
with the effective $T$-dependent masses, 
$\mathcal{M}_{s^i}^2(s^0, \Delta m_p, T) = m_{s^i}^2(s^0, \Delta m_p) + \Pi(T)$, 
where, 
\begin{align}
\Pi(T) = \frac{T^2}{6}\bigl((N_f^2 + 1)f_1 + 2N_f f_2\bigr)\Big|_{f_1 = -f_2 / N_f}\ ,\label{eq:Pi}
\end{align}
is the one-loop self-energy in the infrared limit
in the leading order of the high temperature expansion $\propto T^2$~\cite{Gross:1980br}.
(For a pedagogical review, see~\cite{Quiros:1999jp}).

Using the mass functions given in Eq.~(\ref{eq:massmp}), 
the total effective potential $V_{\rm eff}(s^0, T)$ with the daisy diagrams is given as
\begin{align}
\label{eq:veffT}
V_{\rm eff}(s^0,T) 
=&
\frac{N_f^2-1}{64\pi^2} 
\mathcal{M}_{s^i}^4(s^0, \Delta m_p, T)
\left( \ln{\frac{
\mathcal{M}_{s^i}^2(s^0, \Delta m_p, T)
}{\mu_{_{GW}}^2}} - \frac{3}{2} \right) 
\notag \\
&+\frac{T^4}{2\pi^2} (N_f^2-1) J_B\left(
\mathcal{M}_{s^i}^2(s^0, \Delta m_p, T)
/T^2\right) + C(T)\ .
\end{align}
Here again $C(T)$ is a temperature dependent constant, 
which includes the pseudoscalar contributions.  
We note that $C(T)$ does not affect the phase transition dynamics, 
since the mass functions for pseudoscalars do not depend 
on $s^0$.
In the following, we use this effective potential
to investigate the gravitational wave signal in the electroweak phase transition.

As will be shown later,
the chiral phase transition described by our effective potential
is of very strong first-order with a supercooling follows;
The broken phase nucleation takes place at much lower temperature ($T_n$)
than critical temperature $T_{\text{cr}}$ at which the broken and symmetric vacua equilibrate.
One might be skeptical of using the daisy improved potential for a system at $T = T_n \ll T_{\text{cr}}$
because it partly includes the high-temperature expression of Eq.~(\ref{eq:Pi}).
This problem was addressed in the recent work~\cite{Prokopec:2018tnq}
in a different model of the conformal symmetry breaking, 
and the daisy improved potential was consistent with the potential derived without recourse to
high-temperature assumption at $T\sim T_n$, as long as the high temperature expansion is not performed
for the thermal function $J_B$. Therefore, we assume the applicability of the daisy improvement at $T = T_n$
and provide some discussions in Sec.~\ref{sec:discuss}.

\begin{figure}[!thbp]
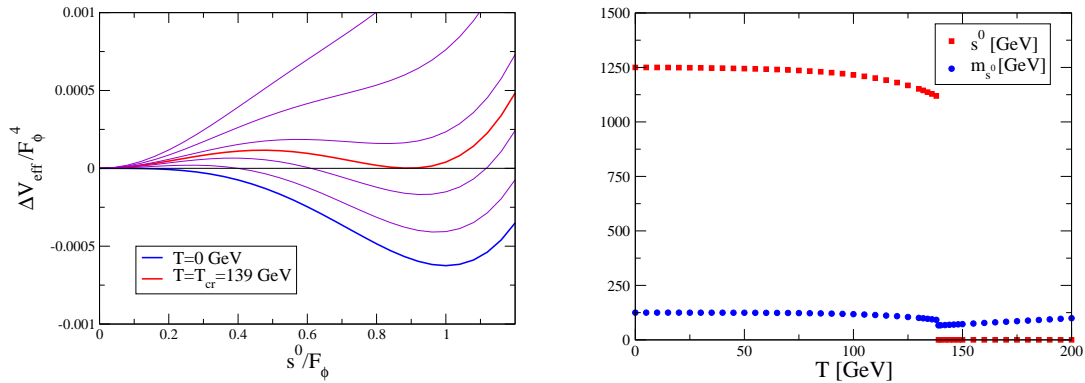

\includegraphics[height=5cm,clip]{./veff_F1250.eps} \quad \quad \quad
\includegraphics[height=5cm,clip]{./s0_F1250.eps}
\caption{
(Left) Effective potential ($\Delta V_{\rm eff} \equiv V_{\rm eff}(s^0,T)-V_{\rm eff}(0,T)$) for various temperature.
The red and blue lines represent the potential at $T = T_{\text{cr}} = 139$ GeV and zero temperature, respectively.
(Right) The vev $\langle s^0 \rangle$ (red squares) and dilaton mass $m_{s^0}$ (blue circles) determined
at the potential minimum as a function of temperature.
}\label{fig:ftptnf8}
\end{figure}

In the left panel of Fig.~\ref{fig:ftptnf8}, we show our effective potential
for various temperature with the benchmark value of the dilaton decay constant: $F_{\phi} = 1.25$ TeV.
The zero temperature potential (blue line) is well-defined (no negative mass in logarithm)
for any value of $s^0$ and flat (zero curvature) near the origin. This is a typical feature of the dilaton-type
potential and contrasted sharply with a usual one loop potential with a negative mass squared at classical level.
The potential monotonically decreases till the minimum and does not possess any barrier at $T = 0$.
This property persists at finite temperature unless it exceeds a small threshold value.
Therefore, in the cooling and expanding universe, the symmetric phase does not survive in the end,
and the cosmological phase transition should complete at certain temperature.
Thus, our model does not suffer from the graceful-exit problem which is recently discussed in the context of
the first-order electroweak phase transition~\cite{Ellis:2018mja}.

With increasing temperature, the potential shape evolves to a double-well structure
and two vacua get equilibrated at $T = T_{\text{cr}} \simeq 139$ GeV (red line)
where the first-order chiral phase transition takes place.
The right panel of Fig.~\ref{fig:ftptnf8} shows the corresponding vev (red squares)
and dilaton mass (blue circles) as a function of temperature.
The strength of the transition is found to be very strong,
$\langle s^0(T_{\text{cr}})\rangle / T_{\text{cr}} \simeq 8.1 \gg 1$,
and strong gravitational waves are created and detectable by LISA project as shown in the next section.
As mentioned in Sec.~\ref{sec:mass},
the dilaton-type potential in Eq.~(\ref{eq:veff}) does not have quadratic negative curvature at the origin
and small finite temperature effects immediately create a shallow potential barrier far separating two vacua.
Therefore, it naturally follows as a dilaton-like potential
that its phase transition becomes strong first-order.

The dilaton mass $m_{s^0}$ (blue points in the right panel) is $125$ GeV at $T = 0$,
and decreases for larger $T$ in the broken phase,
and shows a singular behavior at the critical temperature $T_{\text{cr}}$.
In the symmetric phase, $m_{s^0}$ starts increasing due to the thermal mass effects $\Pi(T)$ given in Eq.~(\ref{eq:Pi}).

\section{Bubble Nucleation and the Gravitational Wave Signal}\label{sec:gw_signal}

In this section, we investigate the spectrum of the stochastic gravitational waves
$h^2\Omega_{\rm{GW}}(f)$, ($h = $ Hubble constant today/100, $f$ = frequency)
which results from the first-order phase transition shown in the previous section.
In principle, $h^2\Omega_{\rm{GW}}(f)$ could be obtained by solving Einstein equation;
The matter sector is specified to the energy momentum tensor for the dilaton
and the plasma velocity fields with the equation of state derived by the effective potential.
However, we do not adopt this strategy but utilize the formulas shown in Appendix~\ref{app:gw};
The $f$ dependence of the $h^2\Omega_{\rm{GW}}$ has been known in the literature
(see Ref.~\cite{Caprini:2015zlo} and references are therein),
and the remaining ambiguities are only three bulk parameters $(T_*, \alpha, \tilde{\beta})$.
Here, $T_*$ denotes the temperature when the gravitational waves are produced.
The parameters $\alpha$ and $\tilde{\beta}$ are associated with the latent heat
and the bubble nucleation rate, respectively.
In the following subsections, we explicitly define and calculate them by using our effective potential,
and compare the obtained gravitational wave spectra with the LISA sensitivity curves.

\subsection{Bubble nucleation with supercooling}\label{subsec:nucl}

The phase transition taking account of the bubble nucleation dynamics in the expanding Universe
is referred to as the cosmological phase transition, which
has been studied in~\cite{Coleman:1977py, Callan:1977pt, Linde:1981zj}.
It is important to calculate the action of the bounce solution of the dilaton field in the early universe,
which characterizes the bubble nucleation through the vacuum tunneling. 
We start from the bubble nucleation probability per unit volume/time ($\Gamma$) given by 
\begin{align}
\label{eq:Gamma}
\Gamma = a(T) e^{-S_E (T)}, 
\end{align}
where $S_E(T)$ is the Euclidean action describing the bubble dynamics at $T$.
At finite temperature, it can be approximately represented as $S_E = S_3/T$, 
where $S_3$ is the three dimensional Euclidean action,
\begin{equation}
\label{eq:S3}
S_3(T) = \int d^3x \Bigl[\frac{1}{2}\bigl(\partial_i s^0(\mathbf{x}, T)\bigr)^2
+ V_{\text{eff}}\bigl(s^0(\mathbf{x}, T), T\bigr)\Bigr]\ .
\end{equation}
The bubble nucleation temperature ($T_n$) is defined as a temperature
at which the nucleation rate $\Gamma$ normalized by the Hubble expansion rate at finite temperature $H(T)$
becomes order 1: $\Gamma(T_n) / H^4(T_n)\sim 1$. Since the prefactor $a(T)$ in Eq.~(\ref{eq:Gamma}) is proportional to $T^4$,
the nucleation condition is rewritten as
\begin{align}
\label{eq:140}
\frac{S_3(T_n)}{T_n} \simeq  4\log\Bigl[\frac{T_n}{H(T_n)} \Bigr]\sim 140\ ,
\end{align}
where,
\begin{align}
\label{eq:hubble}
H(T)^2=\frac{8\pi G_N}{3}\rho_{{\rm rad}}(T)\ ,\qquad
\rho_{{\rm rad}}(T) = \frac{\pi^2 T^4 g_*(T)}{30}\ ,
\end{align}
with $G_N$ and $g_*(T)$ being the Newton constant and the effective degrees of freedom in the plasma, respectively.
We will find $T_n\ll T_{\text{cr}}$ and only SM degrees of freedom are active at $T = T_n$:
$g_*(T_n) = 106.75$.
In Eq.~(\ref{eq:140}), the logarithm term is solely determined by the Newton constant $G_N$ and almost independent of $T$,
and thus approximated as $140$ as indicated by the second equality~\cite{Quiros:1999jp}.

From the fact that the universe is expanding,
the nucleation temperature $T_n$ becomes smaller than the critical temperature $T_{\text{cr}}$
at which the phase transition sets in (two vacua equilibrate). This is particularly true for
the strong first-order phase transition where a strong supercooling holds 
($T_n \ll T_{\text{cr}}$).
After the broken phase nucleation, the latent heat is released
and the suppercooled universe is reheated ($T\to T_{*}$).
Through the above process, the stochastic gravitational waves are generated and red-shifted up until today.
The spectrum is characterized by the reheating temperature $T_*$ (c.f., Eq.~(\ref{eq:T_reh}))
with a red-shift factor and two parameters $(\alpha,\tilde{\beta})$ as shown in Appendix~\ref{app:gw}.
The parameter $\alpha$ represents a normalized latent heat ($\Delta\epsilon(T)$)
determined at the nucleation temperature,
\begin{align}
\label{eq:alpha}
\alpha(T) = \frac{\Delta \epsilon(T)}{\rho_{{\rm rad}}(T)}\ ,\quad
\alpha \equiv \alpha(T_n)\ ,
\end{align}
where the radiation energy density $\rho_{{\rm rad}}$ is defined in Eq.~(\ref{eq:hubble}).
The latent heat $\Delta \epsilon$ is calculated from the effective potential as
\begin{align}
\Delta\epsilon(T) = - \Delta V_{\text{eff}}(T) + T\frac{d}{dT} \Delta V_{\text{eff}}(T)\ ,
\end{align}
where $\Delta V_{\text{eff}}(T)$ is the difference of the effective potential at true and false vacua.
Needless to say, $\Delta \epsilon$ can be defined when two vacua coexists.
A stronger first-order phase transition results in a larger latent heat, and thereby, a stronger gravitational wave signal.

Another parameter $\tilde{\beta}$ is related to the expansion coefficient of $(S_3/T)$
around the bubble nucleation time $t = t_n $, 
\begin{align}
(S_3/T)_t \simeq (S_3/T)t_n -\beta (t-t_n)\ ,\quad
\beta = -\frac{d(S_3/T)}{dt}\Big|_{t = t_n}\ .
\end{align}
By definition, $\beta$ measures how rapidly a bubble nucleates.
In the expanding Universe, the normalized $\beta$,
\begin{align}
\label{eq:tilde_beta}
\tilde{\beta} \equiv \left. \frac{\beta}{H(T_n)}
= T_n \frac{d}{dT}\left(\frac{S_3}{T}\right)\right|_{T=T_n}\ .
\end{align}
is the relevant measure.
The peak frequency of the gravitational wave is proportional to the parameter $\tilde{\beta}$.

In order to obtain $\tilde{\beta}$, one must determine the spatial dilaton field configuration $s^0(\mathbf{x})$
and evaluate the three dimensional effective action $S_3/T$ given by Eq.~(\ref{eq:S3}).
We assume the spherical configuration $s^0(r)$ with $r=\sqrt{x^2+y^2+z^2}$.
The relevant configuration is the bounce solution of the equation of motion derived from Eq.~(\ref{eq:S3}),
\begin{align}
\label{eq:diffeq}
\frac{d^2 s_b^0(r, T)}{dr^2} + \frac{2}{r}\frac{d s_b^0(r,T)}{dr} - \frac{dV_{\text{eff}}(s_b^0,T))}{ds_b^0} = 0\ ,
\end{align}
with the boundary condition of
\begin{align}
\left. \frac{2}{r}\frac{d s_b^0(r)}{dr} \right|_{r=0} = 0\ ,\quad 
s_b^0(r)|_{r=\infty} =0\ .
\end{align}
Here, $r = 0$ correspond to the center of the bubble.
We obtain the bounce solution numerically by using the overshooting/undershooting method.
By tuning $T$, we obtain the bounce action $S_3(T)$ at $T=T_n$,
which satisfies the condition of Eq.~(\ref{eq:140}).

\begin{figure}[!thbp]
\begin{center}
\includegraphics[width=0.4\textwidth]{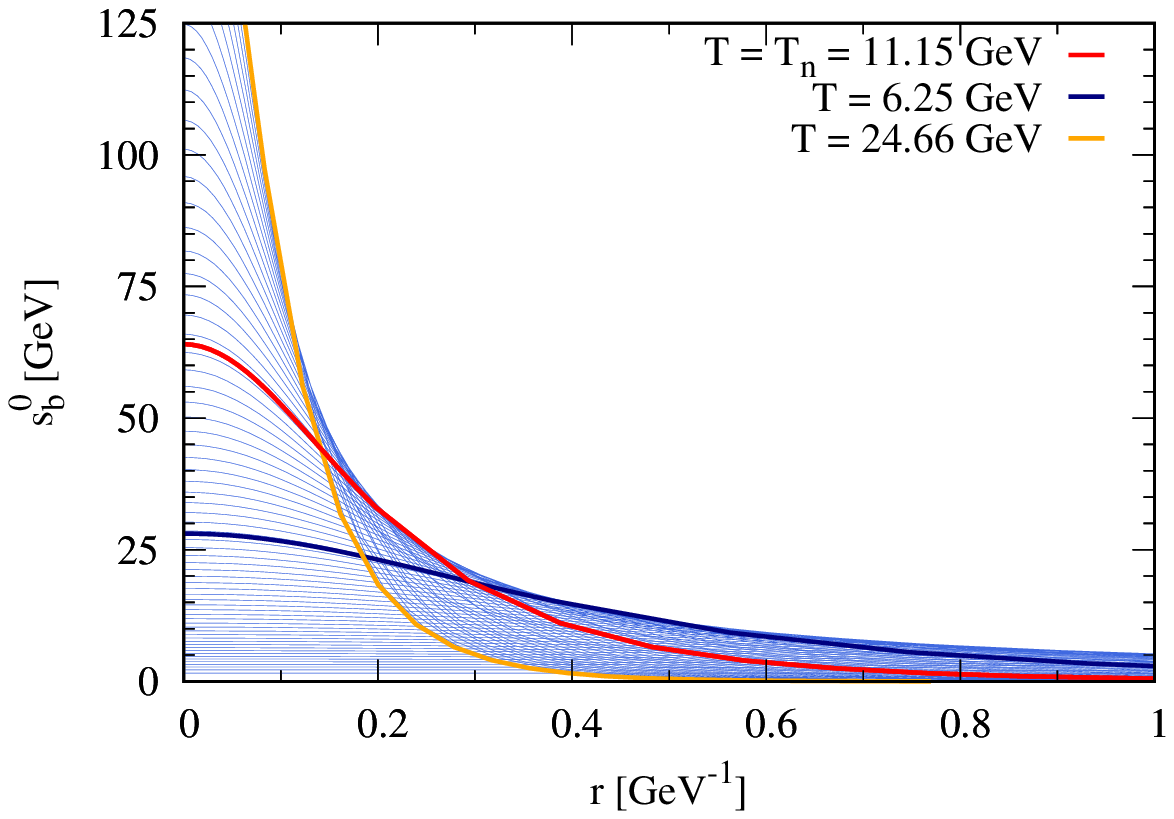}
\includegraphics[width=0.4\textwidth]{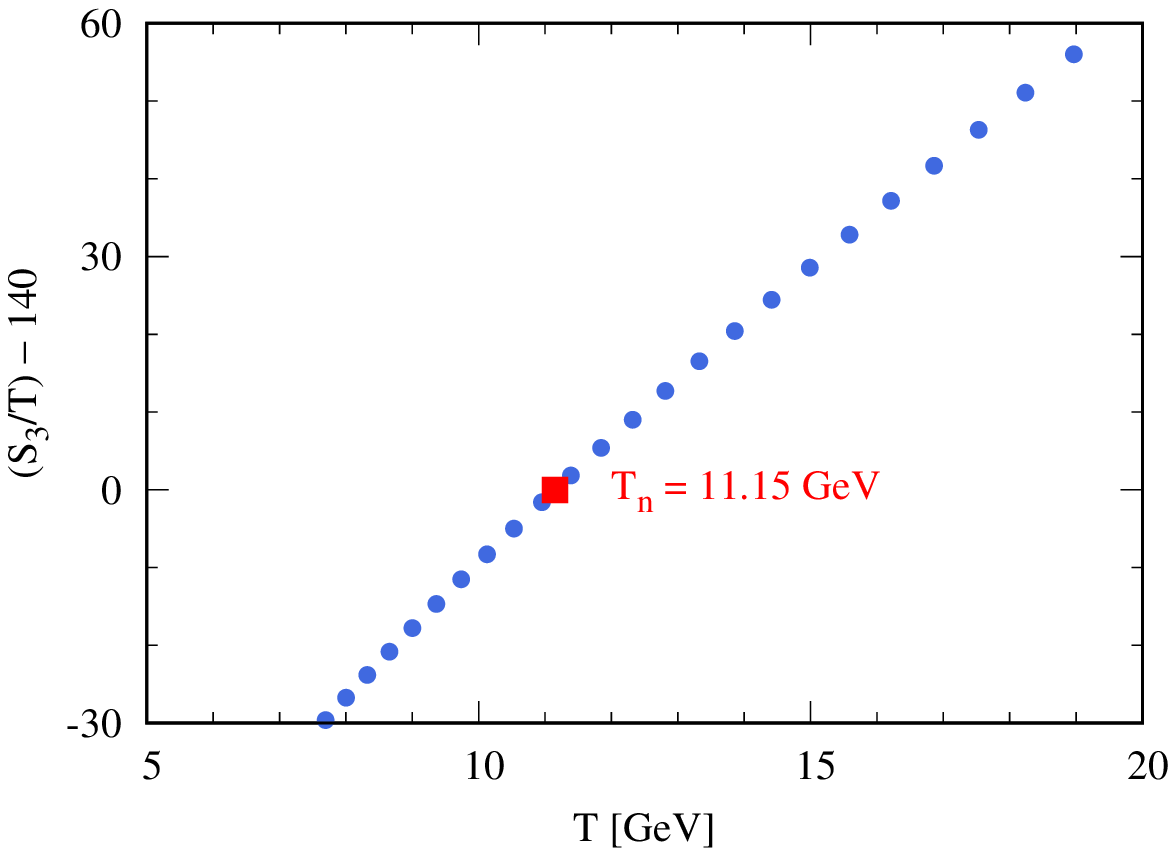}
\end{center}
\caption{
(Left) Bounce solution in Eq.~(\protect\ref{eq:diffeq}). Each line shows a solution at different temperature.
The red line corresponds to the solution which satisfies the nucleation condition Eq.~(\ref{eq:140}).
(Right) The $(S_3/T) - 140$ as a function of temperature $T$. The vanishing point at $T = T_n = 11.15$
is associated with the red line in the left panel.
}\label{fig:phi_bounce}
\end{figure}

In the left panel of Fig.~\ref{fig:phi_bounce},
we plot the bounce solution of the differential equation Eq.~(\protect\ref{eq:diffeq}) for $F_{\phi} = 1250$ GeV.
Each line shows a solution at different temperature.
The orange line is the highest temperature solution in the figure,
and the large portion of $s^0_b(r)$ distributes at small $r$, indicating a small bubble.
As the Universe cools down, the $s^0_b(r)$ gradually distributes at larger $r$.
We find the bounce solution which satisfies the nucleation condition Eq.~(\ref{eq:140}) when
temperature becomes $T = T_n = 11.15$ GeV (red line), at which the gravitational waves are created.
Thus, the strong supercooling $T_n \ll T_{\text{cr}} = 139$ GeV holds.
At smaller temperature, the bounce solution extends to a larger $r$ region as shown by the navy line.
In the right panel, $((S_3/T) - 140)$ is plotted as a function of temperature $T$.
The vanishing point is found at $T = T_n = 11.15$ GeV, where
the slope multiplied by $T_n$ gives the parameter $\tilde{\beta}$ defined in Eq.~(\ref{eq:tilde_beta}).

From the nucleation temperature and the latent heat, we evaluate the reheating temperature as~\cite{Brdar:2018num}
\begin{align}
T_* = T_n(1 + \alpha)^{1/4}\ .\label{eq:T_reh}
\end{align}
In the recent works~\cite{Leitao:2015fmj,Cai:2017tmh,Ellis:2018mja},
it is proposed to determine the $(T_*,\alpha,\tilde{\beta})$
from a percolation temperature $T_p$ instead of the nuclation temperature $T_n$.
It is known that the $T_p$ is comparable or even smaller than the $T_n$.
We expect that the correction by the replacement of $T_n\to T_p$ would be negligible in the gravitational wave spectrum
because of the following consideration. In our model,
there is a threshold temperature $T_{\rm min}$ below which the potential barrier disappears.
This gives a lower bound of the $T_p$. Conservatively, we have investigated the impact of the replacement
of $T_n\to  (1 + \delta)T_{\rm min}$ with $\delta \ll 1$,
and confirmed that the correction to the peak frequency of the gravitational wave signal
is less than 9\% and 7\% for $F_{\phi} = 1000$ GeV and $1250$ GeV, respectively. The correction to the strength
at the peak frequency is negligible.

We summarize the results for $(T_n/F_{\phi}, T_*/F_{\phi}, \alpha, \tilde{\beta})$:
\begin{align}
\label{eq:res_nucl}
&(T_n/F_{\phi},\ T_*/F_{\phi},\ \alpha,\ \tilde{\beta}) = (0.025,\ 0.073,\ 69.5,\ 128.1)\ ,\qquad  (F_{\phi} = 1000\ \text{GeV})\ ,\nonumber\\
&(T_n/F_{\phi},\ T_*/F_{\phi},\ \alpha,\ \tilde{\beta}) = (0.009,\ 0.065,\ 2811.6,\ 87.5)\ ,\qquad (F_{\phi} = 1250\ \text{GeV})\ .
\end{align}
In Ref.~\cite{Caprini:2015zlo}, the benchmark values of ($\alpha,\tilde{\beta}$) are shown for various models.
Most of models give $\alpha \lesssim \mathcal{O}(0.1)$.
In comparison, our $\alpha$ is much larger, particularly for $F_{\phi} = 1250$ GeV,
and thus the strong gravitational wave signal is expected.

\subsection{Numerical results of gravitational wave spectrum}\label{subsec:gw}

In the cosmological first-order phase transition,
the gravitational wave signal $h^2\Omega_{\rm GW}$ is known to be created via the three processes,  
that is, the broken phase bubble collisions, the sound waves in the plasma,
and the magnetohydrodynamics (MHD) turbulence in the plasma.
The corresponding spectra are denoted as $h^2\Omega_{ \phi}$, $h^2\Omega_{\rm sw}$, and $h^2\Omega_{\rm turb}$, respectively,
and the formulas to calculate them are summarized in Appendix~\ref{app:gw}.

The scalar contribution $h^2\Omega_{ \phi}$ is dictated by the bubble wall velocity ($v_w$).
Until the recent development~\cite{Bodeker:2017cim},
it was usually assumed that the bubble walls in the strong first-order transition
can be accelerated without bound and run away with the speed of light
($v_w \to 1$, the runaway bubbles in vacuum)~\cite{Bodeker:2009qy} and the $h^2\Omega_{ \phi}$
could have a significant contribution to the total spectrum $h^2\Omega_{\rm GW}$.
However, more advanced study~\cite{Bodeker:2017cim}
argues that the higher-order corrections to the friction term prevent the bubble wall from becoming runaway,
and the runaway bubbles seem unlikely in the thermal phase transition~\cite{Weir:2017wfa, Jinno:2017fby}.
For the strong phase trantision with a large value of $\alpha > 1$,
the bubble velocity is still approximated by the speed of light $v_w \simeq 1$
but considered as a {\em non-runaway} for which 
the scalar contribution $h^2\Omega_{ \phi}$ is no longer the dominant source of the gravitational wave signals
(see e.g. \cite{Hashino:2018wee, Prokopec:2018tnq}).
We assume that this is the case since we have obtained a large value of $\alpha \gg 1$
with strong supercooing (see Eq.~(\ref{eq:res_nucl})).

The dominant source now becomes the sound wave contributions $h^2\Omega_{\rm sw}$
which arise from the fluid dynamics
\cite{Hindmarsh:2013xza, Hindmarsh:2015qta, Hindmarsh:2017gnf, Giblin:2013kea, Giblin:2014qia}.
As seen in Eq.~(\ref{eq:omega_sw}),
$h^2\Omega_{\rm sw}$ is characterised by the fraction ($\kappa_v$) of the released latent heat
that goes to the plasma bulk motion.
In our case with non-runaway bubbles whose velocity is close to the runaway,
$\kappa_v$ is approximately given as
\begin{align}
\label{eq:kv}
\kappa_v \simeq \alpha (0.73 + 0.083\sqrt{\alpha}+\alpha)^{-1}\ ,
\end{align}
for $\alpha > 1$~\cite{Espinosa:2010hh}.
The gravitational waves for a smaller $v_w$ with $\alpha \lesssim \mathcal{O}(1)$
will also be studied in the next subsection by using the modified expression of $\kappa_v$.

For the turbulence contribution $h^2\Omega_{\rm turb}$,
we use a form modeled by \cite{Caprini:2009yp, Binetruy:2012ze} based on the Kolmogorov-type turbulence~\cite{Kosowsky:2001xp}.
The parameter $\kappa_{\rm turb}$ in Eq.~(\ref{eq:omega_turb}) describes
the fraction of the released latent heat that goes to the turbulent motion of the plasma.
A numerical simulation suggests $\kappa_{\rm turb} \sim \epsilon \kappa_v$
with $\epsilon =0.05-0.1$~\cite{Hindmarsh:2015qta}, and we adopt $\epsilon =0.05$ in our analysis.

\begin{figure}[!thbp]
\begin{center}
\hspace{5mm}
\includegraphics[width=3in]{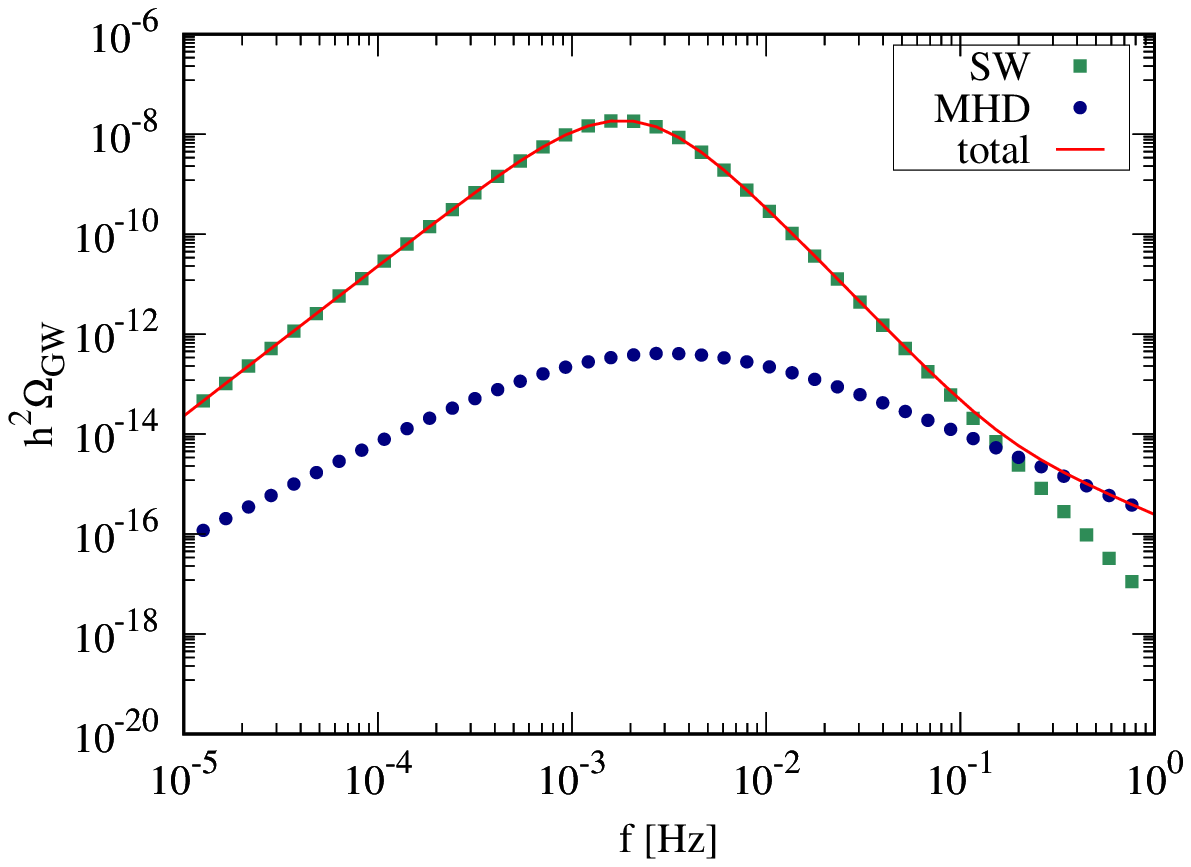}
\includegraphics[width=3in]{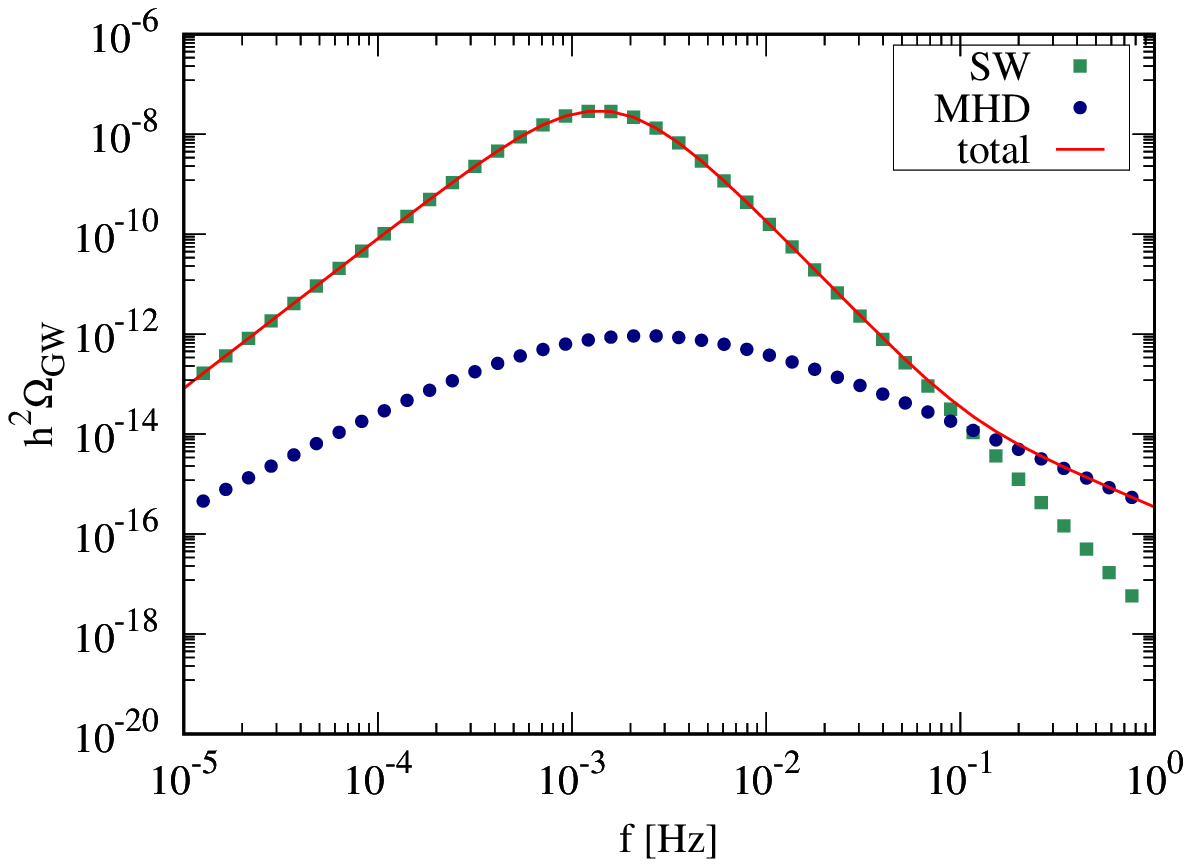}
\end{center}
\caption{
Gravitational wave signal from sound waves (green squares) and MHD turbulence (navy circles).
The red solid line represents the total of them.
Left/Right: $F_{\phi}=1000$ GeV/$F_{\phi}=1250$ GeV.
}\label{fig:gwnf8}
\end{figure}

Fig.~\ref{fig:gwnf8} presents our results for the gravitational wave signals 
from the contributions of the sound waves and MHD turbulence in $N_f=8$ case
for two values of $F_{\phi} =$ 1000 GeV (left) and 1250 GeV (right) favored
in the walking tecnhicolor model~\cite{Matsuzaki:2015sya}.
The peaks locate around $10^{-3} \sim 10^{-2}$ Hz, 
where the dominant contributions come from the sound waves and the strength ($h^2\Omega_{\text{GW}}$) exceeds $10^{-8}$.
The small MHD contributions simply result from using a small value of the fraction $\epsilon = 0.05$.
At large frequency ($f \gg f_{\text{turb}}$), the slope in MHD is determined by the Kolmogorov turbulence model
with the power-law scaling $\sim f^{-5/3}$ while the asymptotic scaling of
the sound wave behaves as $\sim f^{-4}$ ($f\gg f_{\text{sw}}$). As a result, the MHD eventually dominates at large $f$.
The important region comparable to the LISA experiments is $f\sim 10^{-4}$ - $10^{-2}$ Hz,
thus the sound waves plays a crucial role.

\begin{figure}[!thbp]
\begin{center}
\includegraphics[width=3in,clip]{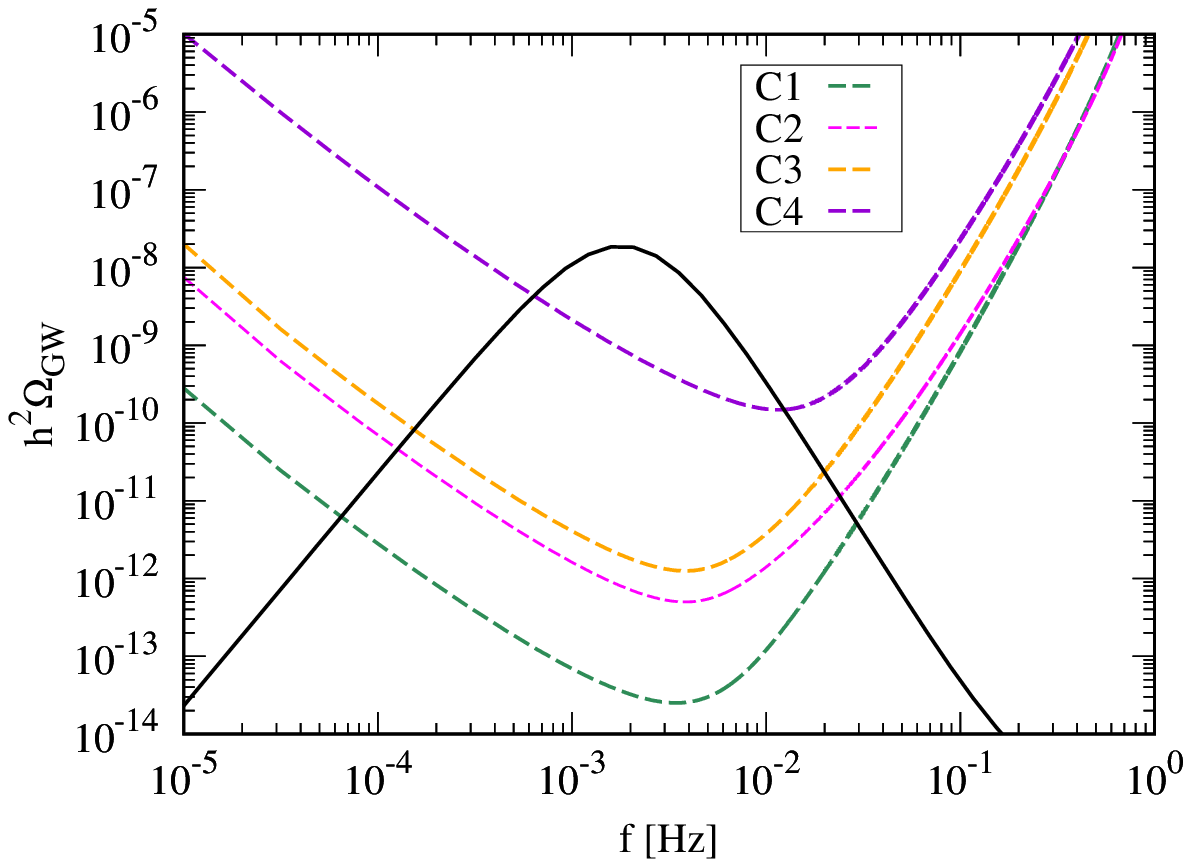}
\includegraphics[width=3in,clip]{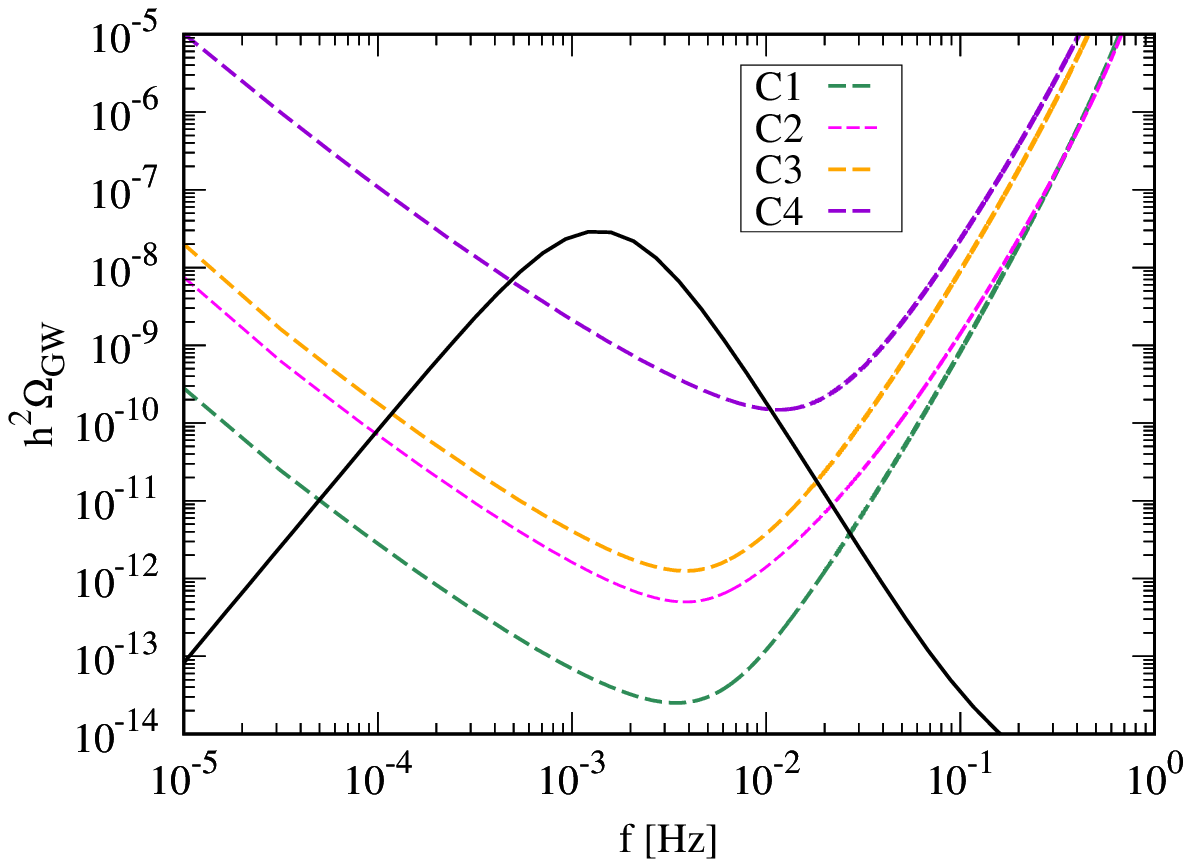}
\end{center}
\caption{
Gravitational wave signal (black solid line) compared with 
LISA sensitivities for configurations C1 - C4 (dashed lines)~\protect\cite{Caprini:2015zlo}.
Left/Right: $F_{\phi}=1000$ GeV/$F_{\phi}=1250$ GeV.
}\label{fig:gw_vs_lisa}
\end{figure}

In Fig.~\ref{fig:gw_vs_lisa}, we show our gravitational wave signals for the total contributions (sound waves + MHD turbulence)
and compare them with LISA sensitivities expected in four representative configurations (C1 - C4)~\cite{Caprini:2015zlo}.
In $F_{\phi} = 1000$ GeV (left panel), our gravitational spectrum 
achieves very strong signals detectable in all (C1 - C4) LISA configurations.
The strong signal is a characteristic feature of the Gildener-Weinberg type potential with a scale symmetry.
In $F_{\phi} = 1250$ GeV (right panel), the strength of the signal gets slightly larger than the $F_{\phi} = 1000$ GeV case
and the peak position is shifted to somewhat smaller frequency region. The signal has a large overlap to LISA sensitivities again.
The peak shift comes from a smaller nucleation temperature with a stronger phase transition at larger $F_{\phi}$.
Interestingly, the particular choices of $F_{\phi}$ motivated by the walking technicolor model result in peaks
close to the most sensitive frequency region in LISA.

\subsection{Discussion}\label{sec:discuss}

\subsubsection{Top quark correction}\label{sec:discuss_top}
We investigate a top quark correction to the gravitational wave spectra shown above.
As explained in Sec.~\ref{sec:dxpt}, the top quark effects
are expected to be subdominant comparing to the flavor non-singlet scalars ($s^i$).
To confirm this reasoning, we add the effective potential from the top quark one-loop contributions,
\begin{align}
V_{\rm eff}^{(t)}(s^0)
= \frac{-12}{64\pi^2}m^4_{t}(s^0)
\left( \ln{\frac{m_{t}^2(s^0)}{\mu_{_{GW}}^2}} - \frac{3}{2} \right) -12\frac{T^4}{2\pi^2}J_F(m^2_t(s^0) / T^2)\ ,
\label{eq:veff_top}
\end{align}
to the original potential of Eq.~(\ref{eq:veffT}), and repeat the calculation of the gravitational wave spectra.
Here, the fermionic thermal function $J_F$ is given by
\begin{align}
\label{eq:JF}
J_F(x)= \int_0^\infty t^2 
\ln{\left(1 + e^{-\sqrt{t^2+x}}\right)} dt\ .
\end{align}

\begin{figure}[!thbp]
\begin{center}
\includegraphics[width=3in,clip]{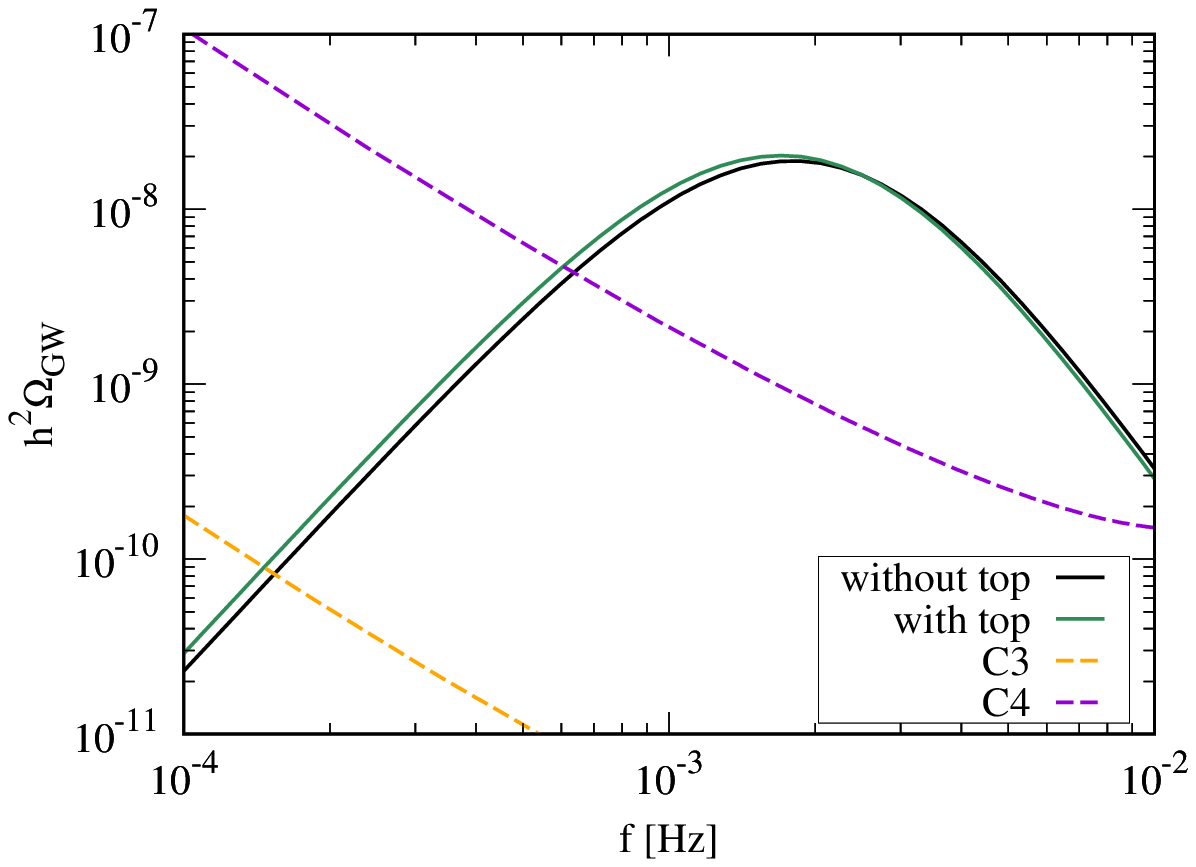}
\includegraphics[width=3in,clip]{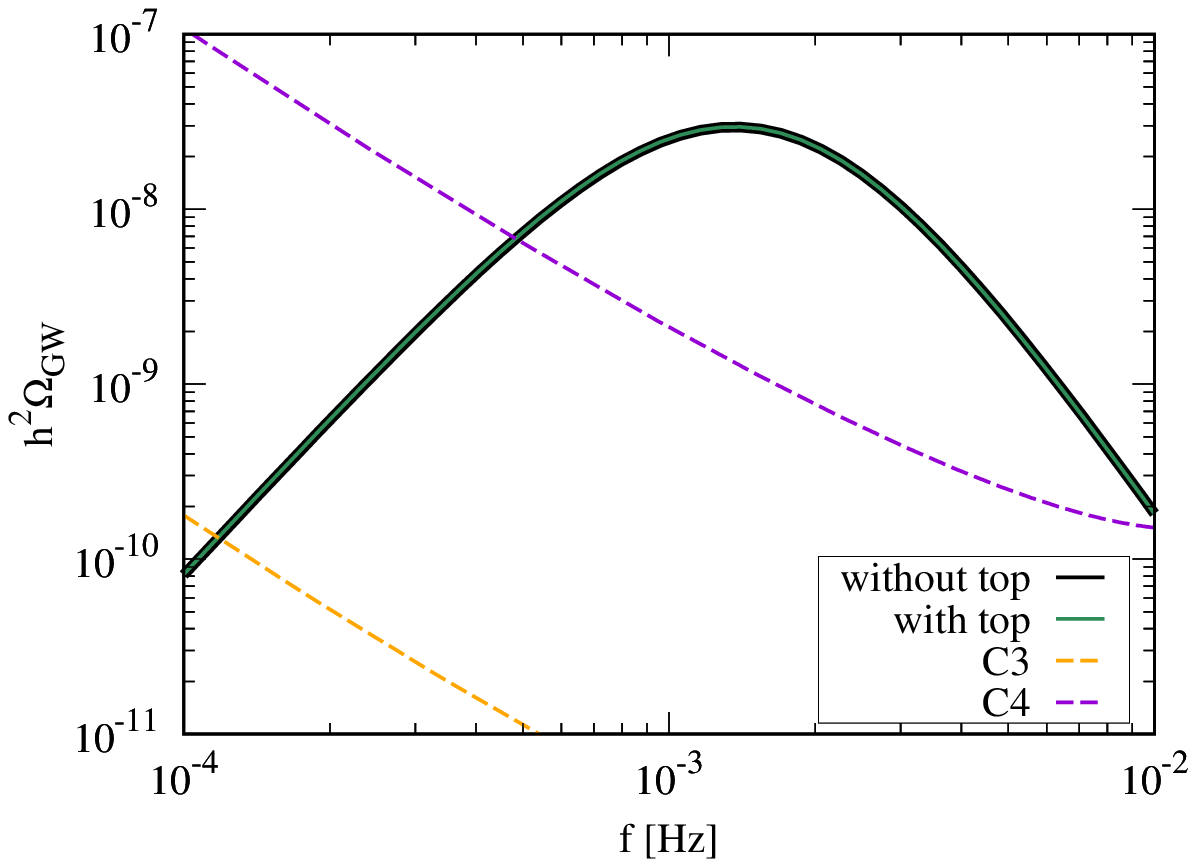}
\end{center}
\caption{
Gravitational wave signal with/without top quark contributions (green/black solid lines).
For comparisons, LISA sensitivities curves for two severe configurations, C3 and C4 (dashed lines),
are also shown~\protect\cite{Caprini:2015zlo}.
To see the top quark effects, the frequency region around the peak is zoomed in.
Left/Right: $F_{\phi}=1000$ GeV/$F_{\phi}=1250$ GeV.
In the right panel, two lines are almost on top of each other.
}\label{fig:gw_top}
\end{figure}

In Fig.~\ref{fig:gw_top}, we compare the gravitational wave spectra
with/without top quark contributions (green/black solid lines).
With top quarks, the peak frequency acquires about 5\% correction 
to the infra-red direction in $F_{\phi} = 1000$ GeV case (left panel).
The modification of the strength at the peak is tiny and negligible.
Comparing to the two severe constraints of LISA sensitivity (C3 and C4),
the top quark correction does not spoil the detectability of the gravitational waves.
The same statement follows in the case of $F_{\phi} = 1250$ GeV (right panel),
where the top contribution is extremely small (0.2\%). 
Since the top correction is suppressed with $m_{\phi} / F_{\phi}$,
the larger $F_{\phi}$ results in the smaller modification.

\subsubsection{
Stability against soft scale symmetry breaking with flavor non-singlet scalar}\label{sec:discuss_ms}

As explained in Sec.~\ref{sec:mass}, the soft breaking mass term Eq.~(\ref{soft}) mimics the explicit breaking
of the $SU(8)_L \times SU(8)_R$ chiral symmetry of the walking technicolor model
and the arbitrary value of the mass parameter $\Delta m_p$  does not affect at all
the strong gravitational wave signals originating from the scale symmetry.
Within the scope of the walking technicolor model, there is no reason to introduce any other mass terms.
Thus, the strong gravitational wave spectra shown in the previous section are the main results
from the walking technicolor scenario.

However, one might concern about some mass terms originating from 
some unknown physics other than the walking technicolor model (with SM and ETC gauging),
such as GUT, gravity, SUSY, and so on. Since our results in the previous section rely on the scale symmetry,
a small mass perturbation, outside of the walking technicolor model, could result in a sizable correction.
In this subsection, we discuss a stability of our walking technicolor results against
such a small mass perturbation 
from outside of the walking technicolor model setting. This might be useful to see a possible high scale physics even beyond the walking technicolor if any. 

 As a specific and simple example,
 we consider the flavor non-singlet scalar $s^i \, (i\ne 0)$ mass term,
\begin{align}
\label{eq:softs}
V_{\rm soft}^{(s)} = \frac{(\Delta m_s)^2}{2} \sum_{i=1}^{N_f^2-1}(s^i)^2\ ,
\end{align}
which gives an extra explicit breaking of the chiral and scale symmetry other than those already discussed, i.e., $s^i$ mass in Eq.(\ref{eq:mass}) due to $\langle s^0\rangle\ne 0$ and $\Delta m_p$, the masses nicely corresponding to those due to
the chiral condensate together with  the SM/ETC gauging in the waking technicolor.
As emphasized above, this 
term has an origin
from
something beyond the walking technicolor. 
Therefore, the 
pseudodilaton ($s^0$), which is the main 
ingredient
in the walking technicolor 
model 
is not included in Eq.~(\ref{eq:softs}). 

As will be shown 
later, the finite $\Delta m_s$ 
in fact 
affects the phase transition dynamics and the gravitational wave spectra
and 
thus are
suitable to investigate the stability of our results.
From the 
purely 
phenomenological 
viewpoint,
a finite $\Delta m_s$ 
extends
the parameter space 
in the mass spectrum 
of a generic model
as  shown in Appendix A.

In the context of the walking technicolor model,
the mass parameter $\Delta m_p\sim \mathcal{O}(1)$ TeV in Sec.~\ref{sec:mass} is a consequence of
the enhancement by the large anomalous dimension $\gamma_m \simeq 1$%
of the technicolor chiral condensate. 
On the contrary, $\Delta m_s$ 
having
no relation to the walking technicolor do not 
acquire
the enhancement associated with the large $\gamma_m$.
Therefore, we assume,
\begin{align}
\Delta m_s \ll \Delta m_p\ .\label{eq:ms_small}
\end{align}
In other words, the mass perturbation comparable or larger than $\Delta m_p$
is different subject from the walking technicolor scenario and beyond the scope of this paper.

Combining Eq.~(\ref{eq:softs}) with the pseudoscalar mass term Eq.~(\ref{soft}),
the $s^0$-dependent mass functions at tree level reads,
\begin{align}
\label{eq:massms}
m_{s^0}^2(s^0, \Delta m_p, \Delta m_s) &= 0, \quad
m_{s^i}^2(s^0, \Delta m_p, \Delta m_s) = (\Delta m_s)^2 + \frac{2f_2}{N_f} (s^0)^2,
\notag \\
m_{p^a}^2(s^0, \Delta m_p, \Delta m_s) &= (\Delta m_p)^2. 
\end{align}
The phase transition dynamics is modified via the shift of
$(2f_2/N_f) (s^0)^2 \to (\Delta m_s)^2 + (2f_2/N_f) (s^0)^2$ in $m_{s^i}^2$ with finite $\Delta m_s$.
The details of the mass spectra at one loop with $(\Delta m_s)^2$ are summarized in Appendix A.
Using Eq.~(\ref{eq:massms}), the total effective potential becomes,
\begin{align}  \label{eq:veffTs}
V_{\rm eff}(s^0, \Delta m_p, \Delta m_s, T)
=&
\frac{N_f^2-1}{64\pi^2} 
\mathcal{M}_{s^i}^4(s^0, \Delta m_p, \Delta m_s, T)
\left( \ln{\frac{
\mathcal{M}_{s^i}^2(s^0, \Delta m_p, \Delta m_s, T)
}{\mu_{_{\rm GW}}^2}} - \frac{3}{2} \right), 
\notag \\
&+\frac{T^4}{2\pi^2} (N_f^2-1) J_B\left(
\mathcal{M}_{s^i}^2(s^0, \Delta m_p, \Delta m_s, T)
/T^2\right) + C(T)\ ,
\end{align}
with,
\begin{align}
\label{eq:m_thermal}
\mathcal{M}_{s^i}^2(s^0, \Delta m_p, \Delta m_s, T)
= m^2_{s^i}(s^0, \Delta m_p, \Delta m_s) + \Pi(T)\ ,
\end{align}
where the thermal mass $\Pi(T)$ is given in Eq.~(\ref{eq:Pi}).
We require that the following properties remain intact for arbitrary $\Delta m_s$;
(1) the vev $\langle s^0\rangle(T = 0)$ determined by the minimum of the potential Eq.~(\ref{eq:veffTs})
is identified with the dilaton decay constant favored by the walking technicolor model, $F_{\phi} = 1.25$ TeV or 1 TeV,
(2) the dilaton mass given by the potential curvature at the vacuum
is identified with the observed SM Higgs mass, $m_{s^0} = 125$ GeV.
Then, the property of the coupling constant $f_2$ is not changed from Eq.~(\ref{benchmark})
while the renormalization scale $\mu_{_{GW}}$ gets modified
as explained in Eq.~(\ref{eq:ren_w_ms}) and the following text.

\begin{figure}[!thbp]
\begin{center}
\includegraphics[width=3in,clip]{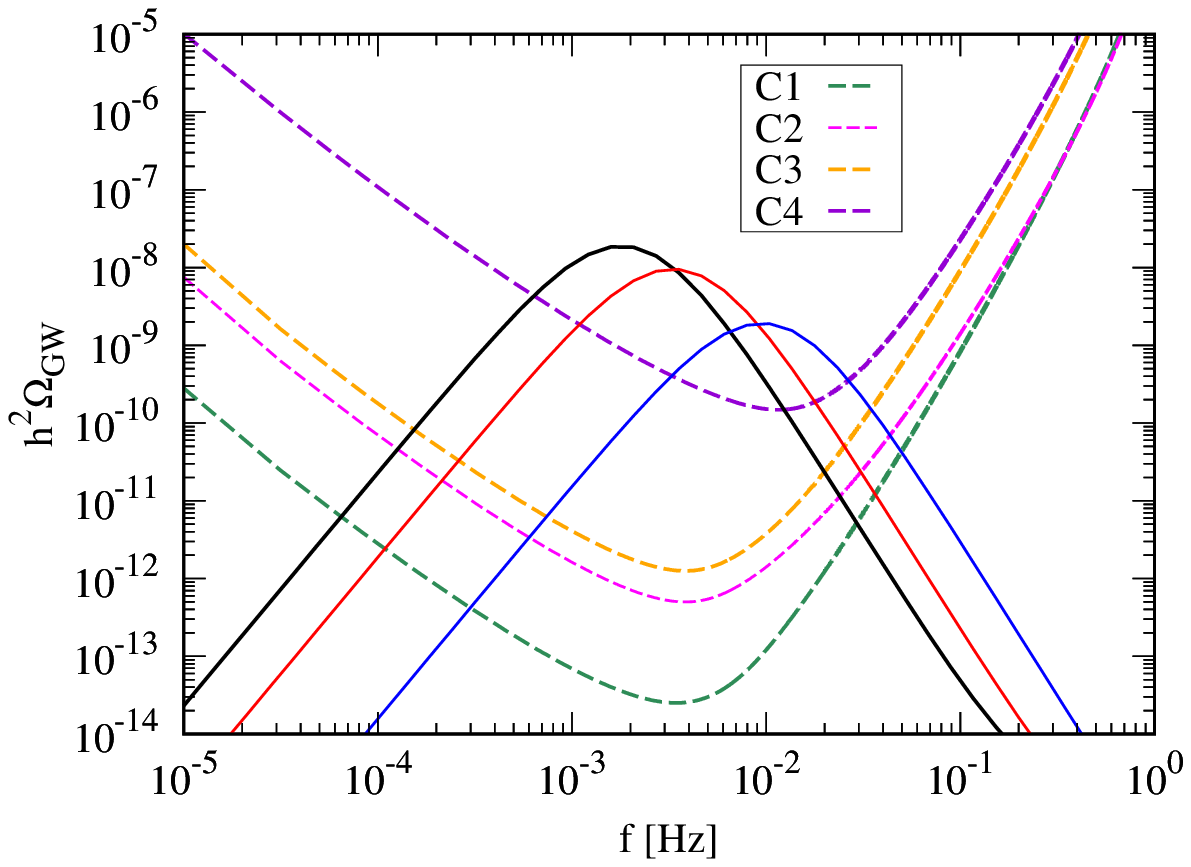}
\includegraphics[width=3in,clip]{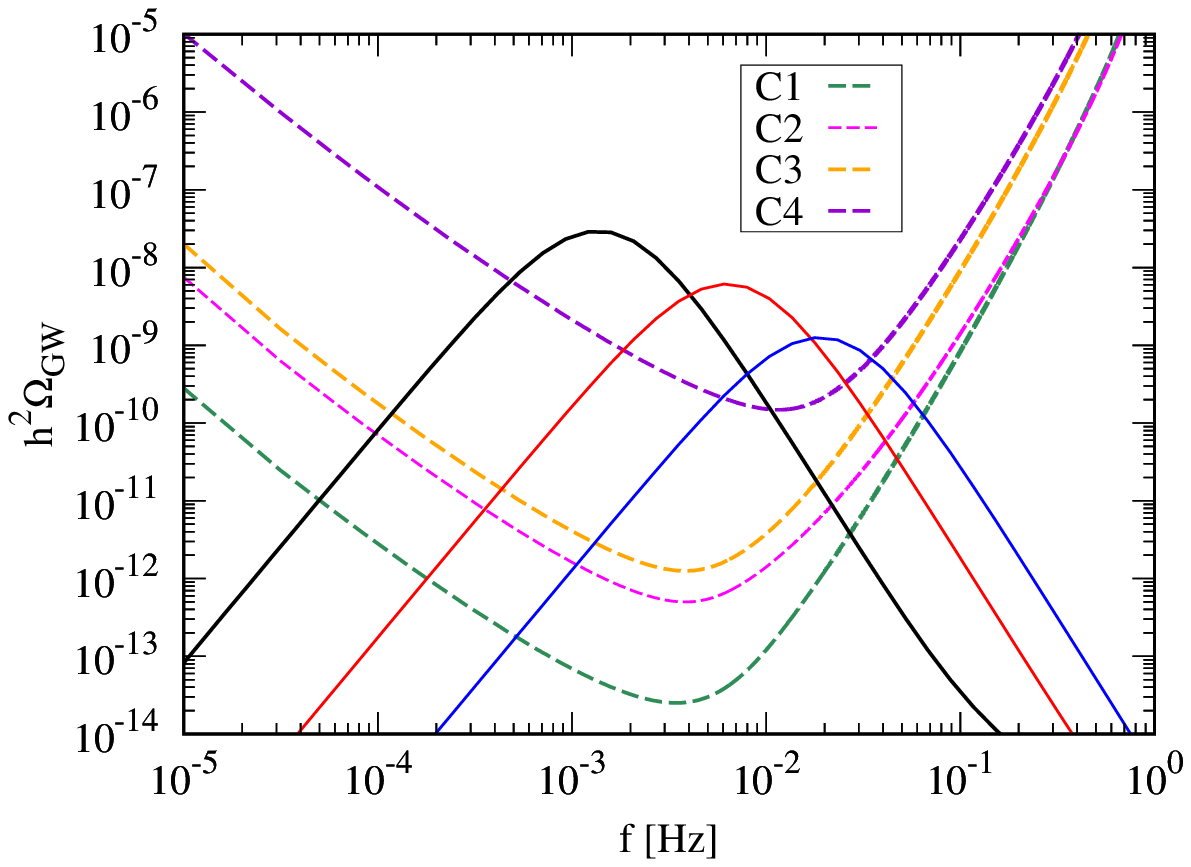}
\end{center}
\caption{
The soft breaking scalar mass ($(\Delta m_s)^2$) effects to the gravitational wave spectra
for $F_{\phi}=1000$ GeV (left) and $F_{\phi}=1250$ GeV (right). 
The black, red, and blue colored solid curves 
represent the results with $(\Delta m_s)^2/F_\phi^2=0$, $0.0001$, and $0.001$, respectively.
The dashed lines represent the LISA sensitivity curves.
}\label{fig:gw_mms}
\end{figure}

In Fig.~\ref{fig:gw_mms}, we show the gravitational wave spectra obtained by using Eq.~(\ref{eq:veffTs})
with $(\Delta m_s)^2 / F_{\phi}^2 = 10^{-4}$ (solid-red) and $(\Delta m_s)^2 / F_{\phi}^2 = 10^{-3}$ (solid-blue),
and compare them to the result with $\Delta m_s=0$ (solid-black) and the LISA sensitivity curves (dashed-lines).
In both $F_{\phi} =$ 1000 GeV (left) and 1250 GeV (right) cases, the peak positions are shifted to the larger frequency region
and the strength becomes somewhat smaller due to the scalar mass perturbations. However, the qualitative feature is not changed
from the case with $\Delta m_s=0$; The signals are still very strong and detectable even in the severe constraint case of C4.
Thus, the strong gravitational wave signal is not spoiled by the small $(\Delta m_s)^2$ perturbation.
This is somewhat surprising because our model relying on the scale symmetry looks sensitive to infinitesimal mass perturbation
but it seems not happening in our model. We can understand this as follows;
The broken phase nucleation takes place at $T_n$ much smaller than the critical temperature $T_{\text{cr}}$,
and the effective potential in the broken vacuum at $T_n$ is close to the zero temperature potential.
Then, in the mass function Eq.~(\ref{eq:m_thermal}),
the dominant scale in the broken vacuum is $\langle s^0\rangle \sim F_{\phi}$ and $(\Delta m_s)^2$ is negligible. 
In the symmetric vacuum, the $s^0$ dependent part disapears in Eq.~(\ref{eq:m_thermal})
but the thermal mass $\Pi(T)$ protects the system from $(\Delta m_s)^2$.
In fact, $(\Delta m_s)^2 < \Pi(T_n)$ is satisfied for small $(\Delta m_s)^2/F_{\phi}^2 = 10^{-4}$ and $10^{-3}$
as shown in Table~\ref{tab:data}. Thus, $(\Delta m_s)^2$ is subdominant in both symmetric and broken vacua.

\begin{table}[!t]
\begin{tabular}{lrrlc}
\hline\hline
$(\Delta m_s)^2/F_{\phi}^2$ & $\Delta m_s$ [GeV] & $T_*$ [GeV] & $f_{\rm peak}$ [Hz]  & $h^2{\Omega}_{\rm GW}|_{\rm peak}$ \\
\hline
0.0           & 0.0   \quad \quad      & 81.2    \quad \quad   & $1.4 \times 10^{-3}$ & $3.0 \times 10^{-8}$ \\
0.0001        & 12.5 \quad \quad       & 81.5   \quad \quad    & $6.3 \times 10^{-3}$ & $6.1 \times 10^{-9}$ \\
0.001         & 39.5  \quad \quad      & 86.4   \quad \quad    & $1.9 \times 10^{-2}$ & $1.3 \times 10^{-9}$ \\
\hline
0.01          & 125.0  \quad \quad     & $\sim$111.3   \quad \quad   & $1.1 \times 10^{-1}$ & $1.4 \times 10^{-11}$ \\
0.02          & 176.8  \quad \quad     & $\sim$137.5   \quad \quad   & $3.4 \times 10^{-1}$ & $1.5 \times 10^{-12}$ \\
0.03          & 216.5  \quad \quad     & $\sim$154.8   \quad \quad   & $6.3 \times 10^{-1}$ & $3.2 \times 10^{-13}$ \\
0.04          & 250.0  \quad \quad     & $\sim$168.0   \quad \quad   & $9.9 \times 10^{-1}$ & $9.3 \times 10^{-14}$ \\
0.05          & 279.5  \quad \quad     & $\sim$178.6   \quad \quad   & $3.7$                & $1.1 \times 10^{-14}$ \\
\hline\hline
\end{tabular}
\caption{Summary table of 
the peak frequencies and the strength of the gravitational wave signals at the peak
as well as the reheating temperature $T_*$ in $F_{\phi} = 1250$ GeV case. For $\Delta m_s^2 / F_{\phi}^2 \geq 0.01$,
we approximated the $T_*$ by the nucleation temperature $T_n$
since the latent heat $\alpha$ becomes small in Eq.~(\protect\ref{eq:T_reh}).
}\label{tab:data}
\end{table}

\begin{figure}[!htbp]
\begin{center}
\includegraphics[width=0.6\textwidth]{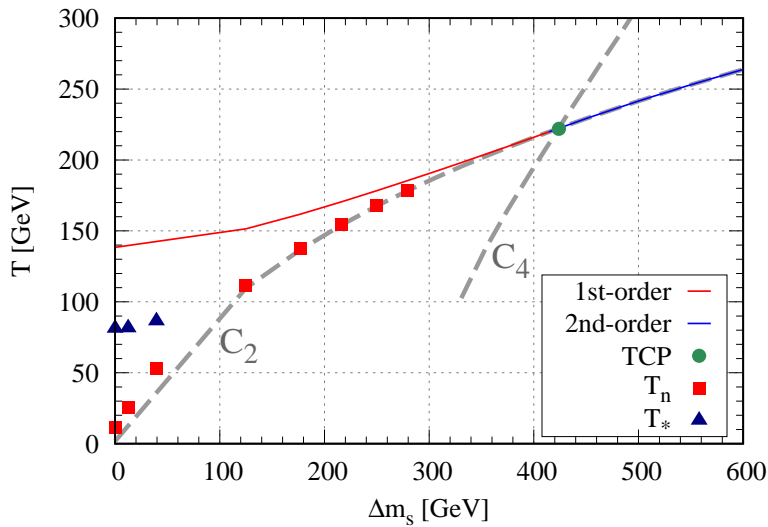}
\end{center}
\caption{
Phase diagram with the chiral phase boundaries and the nucleation and reheating temperature ($T_n$ and $T_*$)
in $T - \Delta m_s$ plane.
}\label{fig:pd}
\end{figure}

We extend our study for a larger mass perturbation $(\Delta m_s)^2$ comparable or larger than the thermal mass $\Pi(T)$.
It is important to know the threshold of $(\Delta m_s)^2$ below which the first-order phase transition survives.
This is the subject of the tri-critical point (TCP) in the the phase diagram in $T - \Delta m_s$ plane,
which we provide in Fig.~\ref{fig:pd} for $F_{\phi} = 1250$ GeV.
It is convenient to consider two auxiliary curves
$C_2$ and $C_4$ which are defined as a pair of $\Delta m_s$ and $T$ satisfying:
\begin{align}
C_2:\quad \frac{\partial^2 V_{\rm eff}(s^0, \Delta m_p, \Delta m_s, T) }{(\partial s^0)^2}\Big|_{s^0\to 0} = 0\ ,\quad
C_4:\quad \frac{\partial^4 V_{\rm eff}(s^0, \Delta m_p, \Delta m_s, T) }{(\partial s^0)^4}\Big|_{s^0\to 0} = 0\ .
\end{align}
The curve $C_2$ is equivalent to the phase boundary in the second-order region (blue solid line)
which appears at large $\Delta m_s$ region.
The crossing point of $C_2$ and $C_4$ corresponds to TCP
at which the first-order phase boundary (red solid line) terminates.
As shown in the phase diagram, the TCP is found at $(\Delta m_{s,{\rm TCP}}, T_{\rm TCP}) = (424, 222)$ GeV in our model.
We select $(\Delta m_s)^2 / F_{\phi}^2 = 0.01 - 0.05 < \Delta m^2_{s,{\rm TCP}} / F_{\phi}^2 \simeq 0.115$
and investigate the gravitational waves in approaching to TCP from first-order transition side.
The nucleation and reheating temperature ($T_n$ and $T_*$) 
should exist between the first-order phase boundary and $C_2$.
As shown by the red squares, 
$T_n$ locates near the $C_2$ and the bubble nucleation is found to happen
just before the potential barrier disappears in the cooling Universe.
The navy triangles in the figure represents the $T_*$ (see Table~\ref{tab:data} for the numerics).
As $\Delta m_s$ becomes larger, the $T_*$ becomes closer to $T_n$ because the latent heat $\alpha$ becomes smaller
in Eq.~(\ref{eq:T_reh}). For $\Delta m_s^2 / F_{\phi}^2 \geq 0.01$,
we approximate the $T_*$ by the nucleation temperature $T_n$
since the $\alpha$ becomes $\mathcal{O}(0.1 - 1)$.

In approaching to TCP, the supercooling strength (distance between $T_n$ and red line) diminishes,
which indicates a gravitational wave signal getting weaker.
However, before reaching at TCP, we find that the spectral peak position
meets the DECIGO target frequencies ($\sim 0.1$ Hz)~\cite{Seto:2001qf,Sato:2017dkf} as shown in Table~\ref{tab:data}.
Depending on the DECIGO configurations,
three sensitivity curves are considered in Ref.~\cite{Kuroyanagi:2014qza}
(one of three curves called ``original'' is found in \cite{Yagi:2011wg}).
The maximal sensitivities allows us to detect the signal of
$h^2\Omega_{\rm GW}\sim \mathcal{O}(10^{-15})$ around  $0.2 - 2.0$ Hz.
For $(\Delta m_s)^2 / F_{\phi}^2 = 0.01 - 0.04$ (except $0.05$),
our results $h^2\Omega_{\rm GW}\sim  \mathcal{O}(10^{-11} - 10^{-13})$
are much larger than at least one of the three sensitivity curves in the corresponding frequency region
and would be detectable in DECIGO.

As explained in the previous section,
we have so far assumed that the wall velocity is close to the speed of light $v_w \simeq 1$.
This assumption is valid for a large latent heat $\alpha(T_*) > \mathcal{O}(1)$ while it becomes $\mathcal{O}(0.1 - 1)$
for $(\Delta m_s / F_{\phi})^2 \geq 0.01$. Therefore, we have checked the gravitational wave spectrum
by using slightly smaller than the speed of sound: $v_w \lesssim v_s = 1/\sqrt{3}$.
In this case, the released latent heat fraction $\kappa_v$ given in Eq.~(\ref{eq:kv})
must be modified as~\cite{Hashino:2016rvx,Espinosa:2010hh}
\begin{align}
\kappa_v \simeq \frac{v_s^{11/5}\kappa_a\kappa_b}{(v_s^{11/5} - v_w^{11/5})\kappa_b + v_s^{6/5}v_w\kappa_a}\ ,
\end{align}
where,
$\kappa_a \simeq 6.9\alpha v_w^{6/5}/(1.36 - 0.037\sqrt{\alpha} + \alpha)$ and
$\kappa_b \simeq \alpha^{2/5}/(0.017 + (0.997 + \alpha)^{2/5})$.
Comparing to the results shown in Table~\ref{tab:data}, the peak frequency $f_{\text{peak}}$ becomes $1.7$ times larger
but still in the DECIGO sensitivity region.
The spectra at $f_{\text{peak}}$ are enhanced by a factor $1.4 - 3.2$ for $(\Delta m_s / F_{\phi})^2 = 0.01 - 0.04$.
Therefore, the detectability in DECIGO remains intact for $v_w \lesssim v_s$ and
does not depend on the details of $v_w$.

Finally, we discuss the limitation of the daisy improved effective potential which we have adopted.
In the recent work~\cite{Prokopec:2018tnq},
it was shown that the daisy improved potential was close to the one derived
by solving a gap equation without high temperature assumptions for the self-energy,
even at $T\sim T_n \ll T_{\text{cr}}$.
This would be understandable as follows;
In the broken vacuum at $T_n \ll T_{\text{cr}}$,
the thermal effects are negligible and whatever the daisy or more sophisticated one does not matter,
and in the symmetric vacuum, nucleation temperature $T_n$ is considered to be {\em high}
in the sense that $T_n$ is larger than mass scales in the symmetric phase,
and the daisy improvement becomes a good approximation there.
However, this is not true anymore in the vicinity of the TCP,
and the results in this work at large $\Delta m_s$ might suffer from this problem.
In this regard, the gravitational waves near TCP should be further investigated
beyond the daisy diagram near future.

\section{Conclusion}

The walking technicolor model, having a large anomalous dimension $\gamma_m\simeq 1$  and a composite Higgs as a pseudo dilaton (``technidilaton'') of the approximate scale symmetry~\cite{Yamawaki:1985zg,Bando:1986bg},
is a viable candidate for physics beyond the Standard Model.
Considering the early Universe at high temperature, the walking technicolor model is expected to undergo
a strong first-order electroweak phase transition 
characteristic to the Coleman-Weinberg (CW) type dilaton potential 
due to the scale symmetric nature. 
Remarkably enough, the spectrum of stochastic gravitational waves $h^2\Omega_{\rm GW}(f)$
from the first-order transition could be observed in LISA and DECIGO experiments in the coming years.
In this regard, we have examined the $h^2\Omega_{\rm GW}(f)$ from the walking technicolor 
by using the linear sigma model with classical scale symmetry which 
simulates the scale symmetric nature of the walking technicolor. As a bench mark model as such, we took the ``one-family model'' with 
$N_f=8$ (four electroweak doublets) in the QCD-like $SU(N_c)$ gauge theory (Lattice studies revealed that $N_f=8,\, N_c=3$ is a walking theory with a light flavor-singlet scalar~\cite{Aoki:2014oha,Aoki:2016wnc,Appelquist:2016viq,Appelquist:2018yqe}
).

Thus our linear sigma model has the chiral $U(N_f)_L\times U(N_f)_R$ symmetry together with the scale symmetry
at classical level particularly for $N_f = 8$ as a bench mark walking technicolor. The scale and chiral symmetries are spontaneously broken at one loop level
through the Coleman-Weinberg (CW) mechanism $\grave{a}$ la Gildener-Weinberg \cite{Gildener:1976ih},
where the scalar in the ray direction (``scalon'') is nothing but 
a light pseudodilaton (technidilaton) 
to be identified with the 125 GeV Higgs particle.
We have shown that our CW potential is equivalent to the most general dilaton potential derived as the effective theory
of the walking technicolor model~\cite{Matsuzaki:2013eva}.
Based on the equivalence, the vacuum expectation value (vev, $\langle s^0\rangle$) is
identified with the dilaton decay constant $F_{\phi}$ rather than the usual linear sigma model
restriction $v_{_{\rm EW}}=\sqrt{N_f / 2}F_{\pi} = 246$ GeV.
For $F_{\phi} = 1.25$ TeV preferred by the walking technicolor model~\cite{Matsuzaki:2015sya},
we have found the coupling constant $f_2^2 \simeq (0.45)^2 \ll 1$ and thus the one loop perturbation theory is justified.
In addition, the radiative corrections from the SM particles can be subdominant and neglected
owing to the large value of $F_{\phi}$.

We have introduced a soft chiral symmetry breaking by the pseudoscalar mass term $(\Delta m_p)^2$ given in Eq.~(\ref{soft}) to 
mock up the NG bosons masses of flavor-non-singlet pseudoscalar ($p^{i = 1 - 63} = \pi^{i}$ = NG-pions)
generated from the gauge interactions of the ETC, 
and the Standard Model $SU(3)\times SU(2)\times U(1)$, and 
$\eta^{\prime}$ mass  by the chiral anomaly.
This also results in 
the mass hierarchy between the light pseudo dilaton (flavor-singlet scalar, $s^0$) and the other massive scalars: 
flavor-non-singlet scalar ($s^{i = 1 - 63} = a_0^i$).
At one loop level this breaking happens not to shift the mass of the pseudo dilaton consistently with the PCDC relation.

In order to investigate the thermal phase transition in our model,
we have considered the one loop thermal effective potential with a daisy diagram improvement.
Due to the scale symmetric feature of our CW potential,
we have observed a very strong first-order electroweak phase transition
which is not affected by the soft mass term given in Eq.~(\ref{soft}) at all.
We have shown that the barrier between the symmetric and the broken vacua disappears
at temperature below a small finite threshold, so that our model does not suffer from the graceful-exit problem
which is recently discussed in the context of the first-order electroweak phase transition~\cite{Ellis:2018mja}.

In order to investigate the gravitational waves from the resultant 
first-order phase transition,
we have numerically examined the bounce solution for the effective action describing the bubble nucleation dynamics.
We have found a strong supercooling with low nucleation temperature ($T_n \ll T_{\text{cr}}$),
large latent heat $\alpha\gg 1$, and the nucleation rate parameter $\tilde{\beta}\sim \mathcal{O}(10^2)$.
By using $T_n$ and $\alpha$, we evaluated the reheating temperature $T_*$.
The results are summarized in Eq.~(\ref{eq:res_nucl}).
(See Sec.~\ref{subsec:nucl} for the parameter definition).

For the given $(T_*,\alpha,\tilde{\beta})$,
the stochastic gravitational wave spectrum $h^2\Omega_{\rm GW}$
is evaluated by using the formulas shown in Appendix~\ref{app:gw}.
For the dilaton decay constant $F_{\phi} = $ 1000 GeV and 1250 GeV,
we have obtained very strong gravitational wave signals $h^2\Omega_{\rm GW}(f)\sim 10^{-8}$
near the best sensitivity frequency region ($f\simeq 10^{-3}$ Hz) of the LISA experiment
and the signals are detectable in all (C1 - C4) representative configurations~\cite{Caprini:2015zlo}.

We have also discussed the effect of the soft breaking mass $(\Delta m_s)^2$ for the flavor non-singlet scalar
given in Eq.~(\ref{eq:softs}). This mass term is not allowed within the scope of the walking technicolor scenario.
The motivation to consider the $(\Delta m_s)^2$ was to assess the stability of our results
against physics other than the walking technicolor model (with SM and ETC gauging), such as GUT, gravity, SUSY, if any.
We have assumed  $\Delta m_s \ll \Delta m_p \sim \mathcal{O}(1)$ TeV because the $\Delta m_s$ does not acquire
the mass enhancement by the large mass anomalous dimension $\gamma_m \simeq 1$ of the technicolor chiral condensate.
For $(\Delta m_s / F_{\phi})^2 = 10^{-4}$ and $10^{-3}$, which are smaller than the thermal mass $\Pi(T)/F_{\phi}^2$,
the gravitational wave signals are still very strong
and detectable in all (C1 - C4) configurations~\cite{Caprini:2015zlo}.
For larger $\Delta m_s$, we have considered the phase diagram in $\Delta m_s - T$ plane for $F_{\phi} = 1250$ GeV,
and located the tri-critical point (TCP: $(\Delta m_{s,{\rm TCP}}, T_{\rm TCP}) = (424, 222)$ GeV).
Approaching to TCP from the first-order region, the gravitational wave signal becomes weaker,
but before reaching at TCP, the spectral peak position
meets the DECIGO target frequencies ($\sim 0.1$ Hz)~\cite{Seto:2001qf,Sato:2017dkf}.
The gravitational wave signals for $\Delta m_s = 125 - 250$ GeV
would be detectable at least one of three sensitivity curves of DECIGO~\cite{Kuroyanagi:2014qza}.
The peak frequencies and strengths are summarized in Table~\ref{tab:data}.

Finally, we notice 
several issues which should be studied in future.
First, it is important to refine the effective potential beyond daisy diagram resummation.
In particular, the difference between the daisy improved and more sophisticated potentials
(e.g. Ref.~\cite{Cornwall:1974vz,Prokopec:2018tnq}) may become significant in the vicinity of the TCP.
Second, it is interesting to refine the estimate of $T_*$ by using the percolation temperature
instead of the nucleation temperature.
Although our results shown in this work might be modified quantitatively by the refinements,
the qualitative feature - very strong gravitational wave signals attributed
to the scale symmetry of walking technicolor model -
would be robust. This suggests that the walking technicolor model
could be probed via the gravitational waves near future in LISA and DECIGO experiments.

\section*{Acknowledgments}
This work is supported in part by JSPS KAKENHI Grants 
No.\,17K14309 and No.\,18H03710.

\appendix
\section{Flavor non-singlet scalar mass}\label{app:ms}

Consider the effective potential in the presence of flavor non-singlet scalar mass $(\Delta m_s)^2$
defined in Eq.~(\ref{eq:softs}) at zero temperature,
\begin{align}
\label{eq:veff0ms}
V_{\rm eff}=\frac{N_f^2-1}{64\pi^2} 
m_{s^i}^4(s^0, \Delta m_p, \Delta m_s) 
\left( \ln{ \frac{m_{s^i}^2(s^0, \Delta m_p, \Delta m_s)}{\mu_{_{GW}}^2}} -\frac{3}{2}
\right) + C,
\end{align}
where $C$ is a constant.
The vev for $s^0$ is obtained from the stationary condition of the effective potential, which 
for non zero value reads
\begin{align}  
\ln{\frac{m_{s^i}^2(\langle s^0 \rangle, \Delta m_s)}{\mu_{_{GW}}^2}} = 1
\quad \Rightarrow \quad 
\langle s^0 \rangle^2 = 
\frac{N_f}{2f_2} 
(e \mu_{_{GW}}^2 - (\Delta m_s)^2)\ .\label{eq:ren_w_ms}
\end{align}  
For arbitrary $\Delta m_s$, the vev is identified with the dilaton decay constant: $\langle s^0\rangle = F_{\phi}$.
From Eq.~(\ref{eq:ren_w_ms}), the renormalization scale is determined as
$\mu_{_{GW}}^2 = e^{-1}\bigl((\Delta m_s)^2 + 2f_2F^2_{\phi}/N_f\bigr)$.

The mass spectra in the presence of $(\Delta m_s)^2$ in terms of $\langle s^0 \rangle$ are given as
\begin{align}
m_{s^0}^2 =& \frac{f_2^2}{2\pi^2} \frac{N_f^2-1}{N_f^2} \langle s^0 \rangle^2 = (125\ \text{GeV})^2\ ,
\notag \\ 
m_{s^i}^2 =& 
(\Delta m_s)^2 +\frac{2f_2}{N_f} \langle s^0 \rangle^2
+ \frac{f_2}{32\pi^2} 
\biggl\{ 
2N_f (\Delta m_p)^2  
\left( \ln{\left(\frac{(\Delta m_p)^2}{\mu_{_{GW}}^2}\right)} -1 \right)
\notag \\
&
\phantom{
(\Delta m_s)^2 +\frac{2f_2}{N_f} \langle s^0 \rangle^2
+ \frac{1}{32\pi^2} \big\{
}
+
f_2 \langle s^0 \rangle^2 \left(
\frac{18}{N_f}   
+ \frac{2}{N_f}  \ln{\left(\frac{(\Delta m_p)^2}{\mu_{_{GW}}^2}\right)} \right)
\sum_{j=1}^{N_f^2-1} (d_{jji})^2 
\biggl\}\ ,
\notag 
\\
m_{p^0}^2 =& (\Delta m_p)^2 
+ \frac{4f_2}{32\pi^2} 
\left(\frac{N_f^2-1}{N_f}\right) (\Delta m_p)^2  
\left( \ln{\left(\frac{(\Delta m_p)^2}{\mu_{_{GW}}^2}\right)} -1 \right)\ ,
\notag \\
m_{p^i}^2 =& (\Delta m_p)^2 
+ \frac{f_2}{32\pi^2} 
\left(2N_f-\frac{4}{N_f}\right) (\Delta m_p)^2
\left( \ln{\left(\frac{(\Delta m_p)^2}{\mu_{_{GW}}^2}\right)} -1 \right).
\end{align}
The relation between the dilaton mass $m^2_{s^0}$ and vev $\langle s^0\rangle^2$
is not modified by $(\Delta m_s)^2$ (nor $(\Delta m_p)^2$),
and the identities given by Eq.~(\ref{benchmark}) with the coupling $f_2\simeq 0.45$ remain intact.

\section{Formulas for gravitational waves spectra from first-order phase transitions}\label{app:gw}

It is known that the stochastic gravitational wave spectrum $h^2 \Omega_{\rm GW}$ created in the first-order phase transition
in the early Universe consists of three signals coming from different sources~\cite{Caprini:2015zlo}:
\begin{align}
\label{eq:gw}
h^2\Omega_{\rm GW} \simeq h^2\Omega_{\phi} + h^2\Omega_{\rm sw} + h^2\Omega_{\rm turb}\ .
\end{align}
The first term $h^2\Omega_{\phi}$ is the gravitational wave signal created by bubble collisions
and associated with a kinetic energy of a scalar (dilaton in this work) field.
The second term $h^2\Omega_{\rm sw}$ represents the signal sourced by sound waves in plasma, 
and the third term $h^2\Omega_{\rm turb}$ accounts for Magnetohydrodynamic (MHD) turbulence
in the plasma forming after the bubbles have collided.

Following \cite{Caprini:2015zlo}, we estimate each contribution as follows,
\begin{align}
h^2\Omega_{\phi}(f) &= 1.67 \times 10^{-5}\ \tilde{\beta}^{-2}
\left( \frac{\kappa_{\phi} \alpha}{1+\alpha} \right)^2  
\left( \frac{100}{g_*} \right)^{1/3} \left(\frac{0.11\,v_w^3}{0.42+v_w^2}\right) S_{\phi}(f) ,
\label{eq:Omenv}
\end{align}
where,
\begin{align}
S_{\phi}(f) &= \frac{3.8 (f/f_{\phi})^{2.8}}{1 + 2.8 (f/f_{\phi})^{3.8}} \notag \\
f_{\phi} &= 16.5 \times 10^{3}\ \tilde{\beta} \left(\frac{0.62}{1.8-0.1v_w+v_w^2} \right)   
\left(\frac{T_*}{100\,{\rm GeV}}\right) \left( \frac{g_*}{100}\right)^{1/6} {\rm Hz}\ ,
\end{align}
for the scalar contribution, and 
\begin{align}
\label{eq:omega_sw}
h^2\Omega_{\rm sw}(f) = 2.65 \times 10^{-6} \
\left( \frac{v_w}{\tilde{\beta}} \right)
\left( \frac{\kappa_v \alpha}{1+\alpha} \right)^2  
\left( \frac{100}{g_*} 
\right)^{1/3} 
S_{\rm sw}(f)
\end{align}
where,
\begin{align}
S_{\rm sw}( f ) &= (f/f_{\rm sw})^{3}
\left(\frac{7}{4 + 3(f/f_{\rm sw})^{2}}  \right)^{7/2} 
\notag \\
f_{\rm sw} &= 1.9 \times 10^{-5} {\rm Hz}\ \left( \frac{\tilde{\beta}}{v_w} \right) 
\left(\frac{T_*}{100\,{\rm GeV}}\right) \left( \frac{g_*}{100} \right)^{1/6}\ ,
\end{align}
for the sound waves, and 
\begin{align}
\label{eq:omega_turb}
h^2\Omega_{\rm turb}(f) = 
3.35 \times 10^{-4} \ 
\left( \frac{v_w}{\tilde{\beta}} \right)
\left(\frac{\kappa_{\rm turb} \alpha}{1+\alpha}\right)^{3/2}
\left( \frac{100}{g_*}\right)^{1/3} 
S_{\rm turb} (f)
\end{align}
where 
\begin{align}
S_{\rm turb} (f) &= \frac{(f/f_{\rm turb})^3}
{\left[ 1 + (f/f_{\rm turb}) \right]^{11 / 3} 
\left(1 + 8 \pi f/h_* \right)}\ ,\\
f_{\rm turb} &= 2.7 \times 10^{-5} {\rm Hz}\ 
\left( \frac{\tilde{\beta}}{v_w} \right) 
\left(\frac{T_*}{100{\rm GeV}}\right)
\left(\frac{g_*}{100} \right)^{1/6}\ ,\\
h_* &= 16.5\times 10^{-6} {\rm Hz}\
\left(\frac{T_*}{100{\rm GeV}}\right)
\left(\frac{g_*}{100} \right)^{1/6}\ ,
\end{align}
for the MHD turbulence.

In the above formulas, the parameters $(T_*,\alpha,\tilde{\beta})$ are
calculated from the effective potential in Sec.~\ref{subsec:nucl}.
The bubble wall velocity $v_w$ is specified to be the speed of light $v_w = 1$ or sound $1/\sqrt{3}$.
The parameter $\kappa_v$ and $\kappa_{\rm turb}$ are explained in Sec.~\ref{subsec:gw}.
The value of the paramete $\kappa_{\phi}$ was not considered in this work
because we assumed the bubble contribution $h^2\Omega_{\phi}$ to be neglegible for non-runaway bubbles
(see Sec.~\ref{subsec:gw}).

\end{document}